\documentclass[aps,prb,twocolumn,superscriptaddress,showpacs]{revtex4-1}
\usepackage{amsmath,amssymb,graphicx,stmaryrd}

\usepackage{ifthen}
\newboolean{PRBVERSION}
\setboolean{PRBVERSION}{false}

\ifthenelse {\boolean{PRBVERSION}}
{
\newcommand{\co}{(color online)\ }
\newcommand{\labelling}{labeling}
\newcommand{\labelled}{labeled}
\newcommand{\centre}{center}
\newcommand{\centred}{centered}
\newcommand{\focussed}{focused}
\newcommand{\grey}{gray}
\newcommand{\greyscale}{grayscale}
\newcommand{\analogue}{analog}
\newcommand{\parameterized}{parametrized}
\newcommand{\neighbour}{neighbor}
\newcommand{\behaviour}{behavior}
\newcommand{\favour}{favor}
}
{
\pdfoutput=1
\usepackage{color}
\definecolor{LinkColor}{rgb}{0.256,0.439,0.588}
\usepackage{hyperref}
\hypersetup{
   pdfauthor={Xiaoming Zhang and KSD Beach},
   pdftitle={Resonating valence bond trial wave functions with both static and dynamically determined Marshall sign structure},
   pdfsubject={frustrated quantum magnetism},
   pdfkeywords={spin hamiltonian,} {J1-J2 model,} {Marshall sign rule,} {resonating valence bonds},
   colorlinks=true,
   citecolor=LinkColor,
   linkcolor=LinkColor,
   urlcolor=LinkColor
   }
\usepackage{times}
\usepackage{mathptmx}
\newcommand{\co}{}
\newcommand{\labelling}{labelling}
\newcommand{\labelled}{labelled}
\newcommand{\centre}{centre}
\newcommand{\centred}{centred}
\newcommand{\focussed}{focussed}
\newcommand{\grey}{grey}
\newcommand{\greyscale}{greyscale}
\newcommand{\analogue}{analogue}
\newcommand{\parameterized}{parameterized}
\newcommand{\neighbour}{neighbour}
\newcommand{\behaviour}{behaviour}
\newcommand{\favour}{favour}
}

\DeclareMathOperator{\sgn}{sgn}

\begin{document}

\title{Resonating valence bond trial wave functions with both\\
static and dynamically determined Marshall sign structure}

\author{Xiaoming Zhang}
\affiliation{Department of Physics, University of Alberta, Edmonton, Alberta, Canada T6G 2E1}
\affiliation{Department of Physics and Astronomy, University of Western Ontario, London, Ontario, Canada N6A 3K7 }
\affiliation{Department of Earth Sciences, University of Western Ontario, London, Ontario, Canada N6A 5B7 }

\author{K. S. D. Beach}
\email[Electronic mail:\ ]{kbeach@ualberta.ca}
\affiliation{Department of Physics, University of Alberta, Edmonton, Alberta, Canada T6G 2E1}

\date{March 7, 2013}

\begin{abstract}
We construct energy-optimized resonating valence bond wave functions as a means to sketch out the
zero-temperature phase diagram of the square-lattice quantum Heisenberg model 
with competing nearest- ($J_1)$ and next-nearest-{\neighbour} ($J_2$) interactions. Our emphasis is not on achieving 
an accurate representation of the magnetically disordered intermediate
phase ({\centred} on a relative coupling $g = J_2/J_1 \sim 1/2$ and whose exact nature is still controversial) 
but on exploring whether and how the Marshall sign structure breaks down in the vicinity
of the phase boundaries. Numerical evaluation of two- and four-spin correlation 
functions is carried out stochastically using a worm algorithm 
that has been modified to operate in either of two modes: one in which the sublattice {\labelling}
is fixed beforehand and another in which the worm manipulates the current {\labelling} so as to 
sample various sign conventions.
Our results suggest that the disordered phase evolves continuously out of the 
$(\pi,\pi)$ N\'{e}el phase and largely inherits its Marshall sign structure; on the other hand, the transition from the
magnetically ordered $(\pi,0)$ phase is strongly first order and involves an abrupt change in the
sign structure and spatial symmetry as the result of a level crossing.
\end{abstract}

\ifthenelse {\boolean{PRBVERSION}}
{}
{}

\maketitle

\section{ \label{SEC:introduction} Introduction }

Simple spin models have contributed significantly to our understanding of quantum magnetism.
They consist of mutually interacting spin-$S$ objects arranged in a lattice and are meant to 
describe the behavior of localized electrons in a crystalline environment. Such models are generally viewed as effective,
low-energy descriptions, descended from their electronic parent models by a process of integrating out 
the gapped charge degrees of freedom.\cite{MacDonald88} 

A tremendous variety of spin interactions can arise. In particular, a ``$t/U$''-style power series from the
strong correlation limit generates (or at least motivates) an increasingly complicated zoo of  multi-spin interaction terms.\cite{Fujimoto05, Lauchli05,Sandvik07,Beach07a,Majumdar12} Nonetheless, we know
that even the leading order term in the expansion, corresponding to Heisenberg models with just two-spin interactions, can display
 highly nontrivial physics if the exchange interactions are sufficiently frustrating.\cite{Diep05,Lacroix11} 
 In that case, the ground state may be a magnetically disordered, spin-rotation-invariant state---either liquid\cite{Anderson73}
 or solid\cite{Affleck88,Read89}---having no classical {\analogue}.

Otherwise, conventional magnetic order (at some ordering vector $\mathbf{Q}$)
is a generic feature of the ground state for Heisenberg models in spatial dimension greater than one.\cite{Misguich02,Hastings04} 
The absence of frustration is connected to three inter-related properties: (i) the existence of a 
bipartite {\labelling} such that all antiferromagnetic interactions connect sites in opposite sublattices, 
(ii) strict adherence to a Marshall sign rule,\cite{Marshall55} and (iii) the possibility of transforming mechanistically to a basis in which all amplitudes of the wave function are real and nonnegative. The last of these is why nonfrustrated models can be easily simulated using quantum Monte Carlo approaches.~\cite{Beard96,Syliuasen02,Evertz03}

For the $S=1/2$ case, all three properties are conceptually unified in the language of \emph{valence bonds}.\cite{Rumer32,Pauling33,Hulthen38, Fazekas74,Beach06}
The collinear, $\mathbf{Q}$-ordered ground state of a nonfrustrated Heisenberg model can be described in a bipartite
valence bond basis\cite{Beach06,Beach08} in which the AB sublattice {\labelling} coincides with the alternating 
pattern laid out by $\mathbf{Q}$ 
and only spins in opposite sublattices are bound into singlet pairs. In terms of such a basis $\mathcal{V}_{\text{AB}} = \{ \lvert v \rangle \}$,
the ground state has an expansion $\lvert \psi \rangle = \sum_v \psi(v) \lvert v \rangle$ in which each amplitude
$\psi(v)$ is real and nonnegative. 
The exact amplitudes can be obtained numerically by projection.\cite{Liang90, Santoro99, Sandvik05,Sandvik10,Banerjee10}

It is also possible to find extremely good approximate values of the form 
$\psi(v) \approx \prod_{[i,j]\in v} h(\mathbf{r}_{ij})$,
where $h(\mathbf{r}) > 0$ is a function of the vector connecting bond endpoints. This 
resonating valence bond (RVB) ansatz, due
to Liang, Doucot, and Anderson, \cite{Liang88} strictly enforces the 
geometric tiling constraint on the singlet bonds but ignores additional
bond-bond correlations.\cite{Lin12}
For a magnetically ordered state, one can show that factorizability into individual bond amplitudes
is the correct assumption.\cite{Beach07b,Hasselmann06} Moreover, for nonfrustrated systems, the amplitudes
exhibit power-law decay, and hence the wave function contains bonds on all length scales.

As a specific and illustrative example, we consider the  square-lattice $J_1$--$J_2$ model for spin half. It
has two nonfrustrated limits.
The model with anitferromagnetic nearest-{\neighbour} interactions only ($J_1 = 1$, $J_2 = 0$) exhibits 
a N\'{e}el ordered ground state whose staggered moment is roughly 60\% of its fully polarized, classical value. The state is almost perfectly captured by an RVB wave function whose bond amplitudes are computed
as 
$h(\mathbf{r}) = \sum_{\mathbf{q}}e^{i\mathbf{q}\cdot\mathbf{r}}\bigl[1-(1-\gamma_{\mathbf{q}}^2)^{1/2}\bigr]/\gamma_{\mathbf{q}}$.
Here, $\gamma_{\mathbf{q}} = (\cos q_x + \cos q_y)/2$, and the wave-vector sum is taken over a Brillouin zone reduced with respect to $\mathbf{Q} = (\pi,\pi)$. The opposite limit, with \emph{next}-nearest-{\neighbour} interactions dominating ($J_1 = 0^+$, $J_2 = 1$), 
is equivalent to two interpenetrating nearest-{\neighbour} Heisenberg antiferromagnets rotated $45^{\circ}$. 
The spin directions in the two otherwise disjoint subsystems lock to each other\cite{Chandra90} provided that $J_1$ is not strictly zero.
In this case, the ground state is equally well described by the RVB wave function, but with the substitution of 
$\gamma_{\mathbf{q}} = \cos q_x \cos q_y$ 
and a Brillouin zone defined modulo $\mathbf{Q} = (\pi,0)$ or $\mathbf{Q} = (0,\pi)$.

What we present in this paper is an attempt to interpolate between these two limits---through the entire range
of relative couplings that are highly frustrated---using the RVB state as a variational wave function. 
Our approach is inspired by Ref.~\onlinecite{Lou07}, but there are several important differences. The first is simply the scale of the calculation: we have simulated a large number of lattice sizes up to $L=32$ on a dense grid of relative coupling values ($g=J_2/J_1$ ranging from 0 to 1 in steps of $\delta g = 0.01$). Second, we do not require that $h(\mathbf{r})$ respect the full $C_4$ symmetry of the square lattice. Rather,
we impose only the $x$- and $y$-axis reflection symmetry, giving the amplitudes an opportunity either to acquire  (over the course of the energy optimization) the full symmetry or to settle into a state that looks different under $90^{\circ}$ rotation. Third, we explore the space of AB sublattice {\labelling}s by which the bipartite valence bond basis is constructed.

As in Ref.~\onlinecite{Lou07}, we make use of an unbiased, stochastic optimization scheme. Changes to the $h(\mathbf{r})$ values 
are made in the downhill direction of the local energy gradient. Step sizes are randomized, and their magnitude decreases on a power-law schedule. We do not attempt to guide the optimization, other than to ensure that none of the bond amplitudes goes negative; nor do we impose any constraints on the variational
parameters based on any prior knowledge (gleaned, e.g., from mean-field theory\cite{Beach07b} or from a master-equation analysis\cite{Beach09}).

We discover the following. At this level of approximation, the $J_1$--$J_2$ model does indeed support a magnetically disordered intermediate phase. But its width is much smaller than expected: the phase boundaries are found to be at $g_{\text{c}1} \doteq 0.54(1)$
and $g_{\text{c}2} \doteq 0.5891(3)$. The transitions are unambiguously second- and first-order, respectively, with the ground state  achieving the full $C_4$ symmetry for all $g < g_{\text{c}2}$. As the system
is tuned up from $g=0$, increasing frustration eventually extinguishes the $(\pi,\pi)$ ordered moment at $g_{\text{c}1}$ in a continuous fashion. 

The disappearance of magnetic order is preceded by a failure of the Marshall sign rule at $g_{\text{M}1} \doteq 0.398(4)$,
in agreement with the scenario first outlined by Richter and co-workers.\cite{Richter94}
 Still, even though the rule is not strictly obeyed beyond $g_{\text{M}1}$, the Marshall structure inherited from the $g=0$ model remains largely intact throughout the intermediate phase. This is true in the sense that continuing to define the bipartite bond basis from a \emph{checkerboard} sublattice decomposition produces only a microscopic number of negative $h(\mathbf{r})$ values---only $h(\pm 1,\pm 2)$ and $h(\pm 2,\pm 1)$ initially. Moreover, when we allow the AB pattern to arise on its own within the simulation (described in detail in Secs.~\ref{SECT:dynamic} and \ref{SECT:results}), the checkerboard pattern is the one selected whenever $g < g_{\text{c}2}$.

On the other hand, the RVB state at large $g$ explicitly breaks the $90^\circ$ rotation symmetry and has a Marshall sign structure based on a \emph{stripe} sublattice decomposition. As the coupling is tuned down from the $g=\infty$ limit, the $(\pi,0)$ ordered moment is not strongly affected, and it persists with only weak variation (never dropping below 47\% of its fully polarized value) down to $g_{\text{c}2}$, where the spatially symmetric, checkerboard-based RVB wave function takes over as the lowest energy state.
This state in the region $g_{\text{c}1} < g < g_{\text{c}2}$ is, as far as we can tell, featureless. It exhibits no long-range spin or 
dimer order, and it breaks no symmetries. It is not, however, a ``short-range RVB state'' in the usual sense, since it is not made up of predominantly short bonds. Its amplitude function $h(\mathbf{r})$ is highly anisotropic (as anticipated elsewhere\cite{Beach09}) and remains long ranged along the principal spatial axes. Spin correlations appear to be critical and to display circular 
symmetry at long distances, despite the anisotropy of the bond weights. Dimer correlations
decay either exponentially or with a high power law. This is in stark contrast to the usual short-bond-only RVB state,
often referred to as the nearest-{\neighbour} RVB (NNRVB), 
which has spin correlations that decay exponentially~\cite{Liang88}
and dimer correlations that decay algebraically.\cite{Albuquerque10,Tang11}
Moreover, the presence of long bonds implies an absence of the 
topological order\cite{Albuquerque10,Tang11} that is characteristic of a purely short-range RVB state in
two dimensions.

\section{ Model and method }

\subsection{ Frustrated Hamiltonian }
The spin-half, square-lattice Heisenberg model with frustrating interactions has a Hamiltonian
\begin{equation}
H = J_1\sum_{\langle i,j \rangle}\mathbf{S}_i\cdot \mathbf{S}_j+J_2\sum_{\langle \langle i,j \rangle \rangle}\mathbf{S}_i\cdot \mathbf{S}_j,
\end{equation}
where $J_1>0$ and $J_2>0$ are the antiferromagnetic exchange couplings. The summations range over pairs
of adjacent sites $\langle i,j \rangle$ and over farther
pairs $\langle \langle i,j \rangle \rangle$ that sit diagonally across a plaquette.
The ratio $g=J_2/J_1$ is the
key tuning parameter at zero temperature. 
In the classical version of this model ($S\to \infty$), two magnetic phases meet at exactly $g=0.5$, separated by a first-order transition.\cite{Moreo90,Chubukov91,Ferrer93,Ceccatto93} 

In the $S=1/2$ problem, the two magnetically ordered ground states obtain for values $g \lesssim0.4$ and $g\gtrsim 0.6$,\cite{Dagotto89,Schulz96,Oitma96,Bishop98,Singh99,Sushkov01} and a 
magnetically disordered phase intervenes. (There is, however, a good deal of disagreement 
over the exact positions of the critical points; cf.\ Refs.~\onlinecite{Richter10} and \onlinecite{Reuther10},
which put the lower critical point as low as 0.35 and as high as 0.45.)
 The physics of the phase in the intermediate region is not known with complete certainty, but it is commonly believed to be short ranged and not to exhibit any kind of conventional magnetic order. One possibility is a crystalline arrangement of valence bonds, a state with broken translational symmetry in which singlet formation {\favour}s an enlargement of the unit cell beyond that of the underlying square lattice.\cite{Gelfand89,Gelfand90,Singh90,Zhitomirsky96,Leung96,Kotov99,Kotov00,Capriotti00,Takao03,Mambrini06,Murg09,Reuther10,Reuther11,Yu12}
 A featureless spin liquid that does not break any symmetries is another possibility.\cite{Chandra88,Figueirido90,Oguchi90,Locher90,Schulz92,Zhong93,Zhang03,Capriotti01,Capriotti03,Captiotti04a,Capriotti04b}
 
The case for a spin liquid ground state has been advanced by recent tensor product\cite{Wang11} and density matrix renormalization 
group (DMRG)\cite{Jiang11} calculations and by a variational approach based on the entangled-plaquette ansatz.\cite{Mezzacapo12}
With regard to the DMRG result, Sandvik has suggested that the use of a cylindrical geometry complicates the detection of crystalline order.\cite{Sandvik12} His numerical experiments seem to indicate that the mixture of open and closed boundary conditions significantly raises the crossover length scale $\xi$ beyond which bond order takes hold (i.e., where the finite size scaling {\behaviour} of the dimer-dimer correlations is truly in the asymptotic regime). Such questions are difficult to resolve. Unlike in three-dimensional systems, where crystalline bond order, if it is present, is almost always strong,\cite{Beach07a,Block12} in two dimensions it is quite delicate and can easily be disguised by a $U(1)$ effective symmetry for system sizes $L \lesssim \xi$.  (See Sects.~III and IV of Ref.~\onlinecite{Kaul12} and references therein.) Here, we attempt to make the best of this unsatisfactory state of affairs. We simply take the point of view that, 
for the lattice sizes (up to $L=32$) we can simulate, the liquid and the weakly ordered bond crystal are indistinguishable.
  
\subsection{RVB trial wave function}
\begin{figure}
\begin{center}
\ifthenelse {\boolean{PRBVERSION}}
{\includegraphics{patterns.eps}}
{\includegraphics{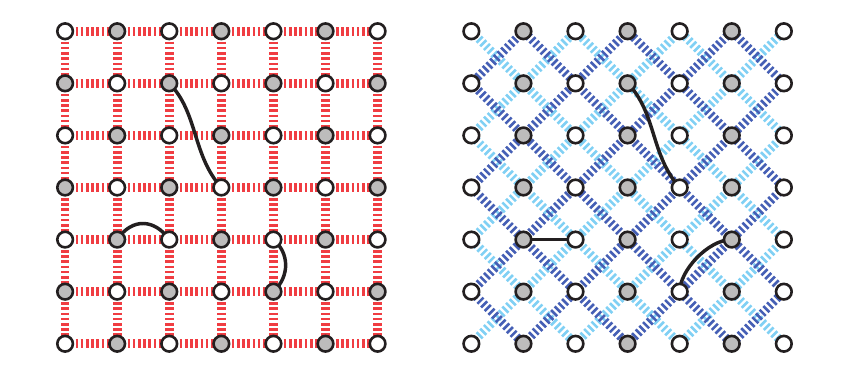}}
\end{center}
\caption{\label{FIG:patterns}\co 
Square grid of lattice sites (circles) whose shading indicates the sublattice membership.
Dashed lines mark the $J_1$~(red) and $J_2$~(blue) exchange couplings.
The basis contains only product states of singlets connecting sites in opposite sublattices.
(Left)~In the limit $g=J_2/J_1 = 0$, a
checkerboard pattern of A and B labels that coincides with $(\pi,\pi)$ magnetic order.
(Right)~In the limit $g=\infty$, a 
stripe pattern that coincides with $(\pi,0)$ order.
In each case, three permissible singlet pairings are indicated.
}
\end{figure}

In quantum Heisenberg models, competing interactions that frustrate the order have the potential
to stabilize exotic quantum phases, but they also render the problem computationally intractable on large lattices. 
Frustrating interactions of even infinitesimal strength cause a sign problem\cite{Loh90} that makes quantum Monte Carlo calculations unfeasible. Moreover, the size of the Hilbert space grows exponentially with system size and is thus beyond the capability of exact diagonalization calculations if we want to get near the thermodynamic limit.
(The record for spin half has recently jumped from 42 sites\cite{Richter04,Richter10,Nakano11,Lauchli11} to 48 sites,\cite{Lauchli12}
a terribly impressive technical feat that nonetheless  limits us to two-dimensional length scales $\sim\!\sqrt{48}$ that are quite small.)
An approximate method based on good trial wave functions is therefore one of the few remaining possibilities
for large systems.

We consider a lattice of $2N$ spins and a factorizable RVB wave function of the form
\begin{equation}
\lvert \psi \rangle = \sum_v \prod_{[i,j] \in v} h(\mathbf{r}_{ij}) \lvert v \rangle,
\end{equation}
where the sum is over all partitions of the lattice into $N$ directed pairs
$v = ( [i_1,j_1], [i_2,j_2], \ldots, [i_N,j_N] )$. To every such \emph{dimer covering} $v$,
there is a corresponding singlet product state; e.g., 
\begin{equation}
\lvert v \rangle = \frac{1}{2^{N/2}}\bigotimes_{[i,j] \in v} \Bigl( \lvert \uparrow_{i} \downarrow_{j} \rangle
- \lvert \downarrow_{i} \uparrow_{j} \rangle \Bigr)
\end{equation}
in the $S=1/2$ case.
The set $\mathcal{V} = \{ \lvert v \rangle \}$ of all possible singlet product states
is both overcomplete and nonorthogonal and constitutes
the so-called valence bond basis.

We can now break up the lattice into two sublattices---groups of sites {\labelled} A and B, equal in number---and
restrict ourselves to a reduced basis in which valence bonds connect only sites in opposite sublattices
(i.e., $v \in \mathcal{V}_{\text{AB}} \simeq S_N$,
rather than $v \in \mathcal{V} \simeq S_{2N}/Z_2^N$).
We adopt the convention that each bond $[i,j]$
is arranged with site $i$ in sublattice A and site $j$ in sublattice B.
This has the advantage of rendering the overlap strictly positive:
$\langle v | v' \rangle = 2^{N_l(C)- N}$, where
$N_l(C)$ is the number of loops in the double
dimer covering $C = (v,v')$.
(In this ``bosonic'' convention, the singlets are AB directed bonds. In the complementary
``fermionic'' convention, the bonds are directionless and all signs are moved into
the overlaps.\cite{Anderson87,Capriotti01,Yunoki04,Cano10,Li12})

To start, we consider two 
families of trial state, each built using a bipartite bond basis consistent with one of 
two static choices of sublattice {\labelling}, viz., the checkerboard and stripe patterns
shown in Fig.~\ref{FIG:patterns}.
Later in the paper, we  go on to describe a procedure in which the trial state
is built using an unrestricted bond basis and the sublattice {\labelling} (and hence the Marshall
sign convention) is determined dynamically. 

The RVB wave function is quite expressive. 
Its degrees of freedom are the full set of $h(\mathbf{r})$ values
with the bond vector $\mathbf{r}$ spanning all lengths and orientations that can be achieved on an $L \times L$ cluster with periodic boundary conditions and that are unique up to whatever symmetries are enforced. (Still, the total number of 
parameters grows only linearly with the number of spins, which is radically slower than the number
of states in the total spin singlet sector.)
Previous calculations of this kind\cite{Lou07,Beach09} considered only the
checkerboard AB pattern and imposed on $h(\mathbf{r}) = h(x,y)$ the full symmetry of the lattice,
such that $h(x,y) = h(|x|,|y|) = h(|y|,|x|)$. In this calculation, we impose
a less restrictive condition, $h(x,y) = h(|x|,|y|)$, that respects reflection symmetry across the lines $x=0$ and $y=0$
but not across the lines $y=\pm x$. For the checkerboard pattern, the number of free parameters is
$(L/2-\eta)(L/2-1)$, where $\eta = (L/2 \mod 2)$ distinguishes between $L/2$ even and odd. 
For the stripe pattern, the count is only slightly higher: $(L/2+1)(L/2-1) = L^2/4-1$.

To recapitulate, our work involves a basis choice. We do not construct the trial wave functions from the largest possible set of valence bond states in which the spins are joined in all possible ways. Instead, we obtain a more restricted basis by dividing the system into two groups of sites (A and B) and keeping only states in which bonds connect A sites and B sites (bipartite bonds). No approximation is involved in this basis choice since the restricted basis is so massively overcomplete that even this subset still spans the relevant part of the Hilbert space. 

But in assigning A and B labels to the sites, we are making a choice about the form of the trial wave function. By working with the checkerboard  and stripe AB patterns, we are in essence adapting the trial wave function to $g = 0$ and $\infty$, respectively,
and taking advantage of the Marshall sign rules that exist in those two limits. We are not biasing the wave function, however, at least 
not in the sense that we are building in magnetic order. The wave functions constructed from either AB pattern are fully capable of 
representing nonmagnetic states.

\subsection{Sampling algorithm}

\begin{figure*}
\ifthenelse {\boolean{PRBVERSION}}
{\includegraphics[scale=0.95]{vbworm.eps}}
{\includegraphics[scale=0.95]{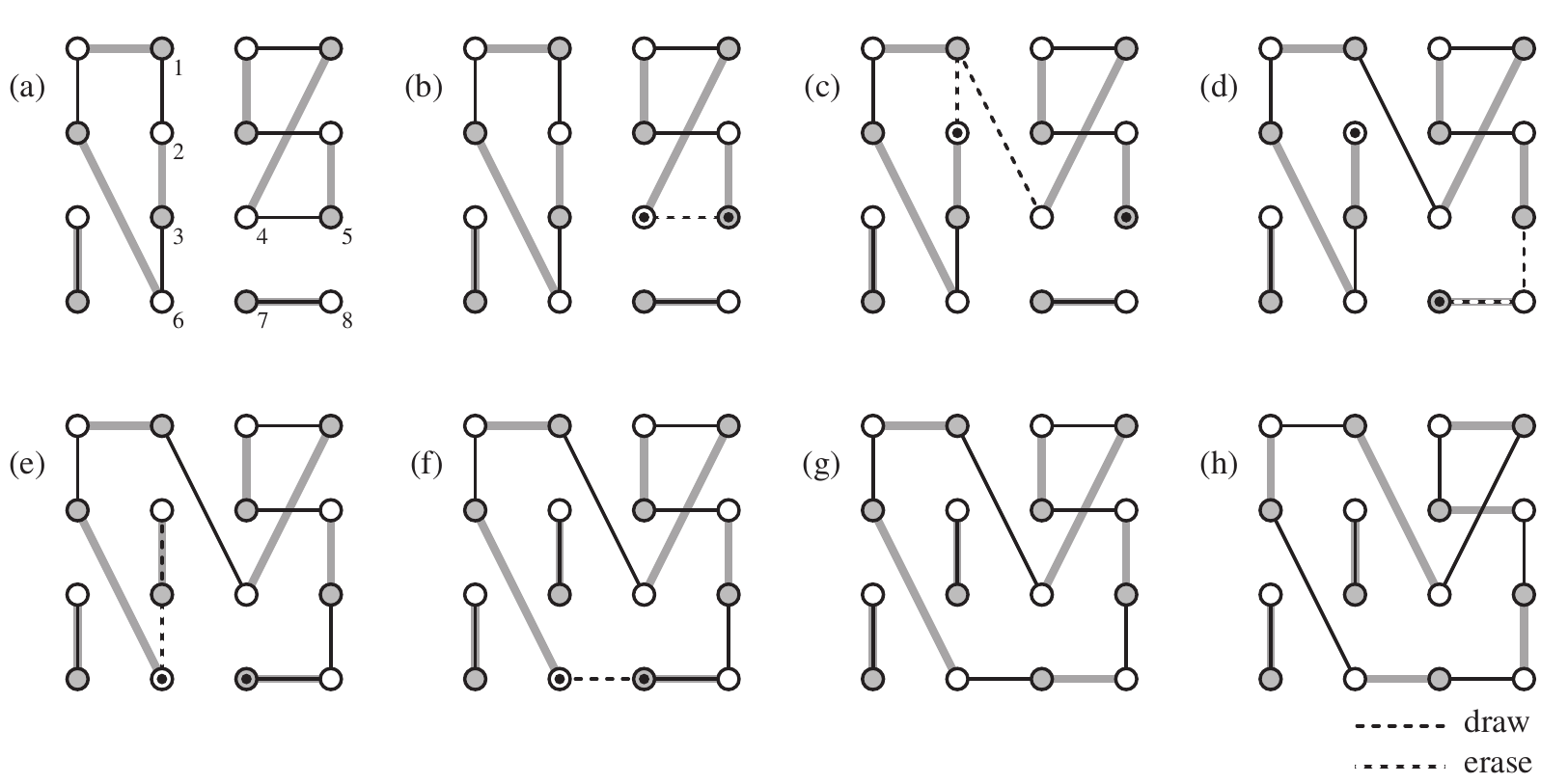}}
\caption{\label{FIG:updates}
(a)~Two superimposed valence bond configurations form a collection of closed loops.
Eight of the sites are numbered for use in Eq.~\eqref{EQ:step5}.
(b)~Breaking one bond leaves an open string with a head and tail located at the former bond's endpoints.
(c)--(e)~The head and tail move by drawing a new bond and erasing the preexisting bond 
emerging from the destination site.
(f)~The open string is closed when the head and tail reconnect.
(g)~The repaired loop configuration.
(h)~Exchanging the background and foreground links in any loop is also a valid
update.
}
\end{figure*}

Every measurement $\langle \hat{O} \rangle = \langle \psi | \hat{O} | \psi \rangle / \langle \psi  | \psi \rangle$
is equivalent to $\langle\!\langle O \rangle\!\rangle$, an ensemble average of the appropriate
 estimator $O:~\!C \to O(C)$ in the gas of fluctuating loops described by
\begin{equation} \label{EQ:Z0}
Z = \frac{1}{q^N} \sum_{C} q^{N_l(C)} \prod_{[i,j] \in C} h_{ij}.
\end{equation}
As before, $C = (v,v')$
is a loop configuration arising from the superposition of two dimer coverings,
and $N_l(C)$ counts the number of loops. The value $q=2$ is the loop fugacity appropriate for $S=1/2$.
When Marshall's theorem holds, the bond amplitudes satisfy $h_{ij} \ge 0$
and thus every term in Eq.~\eqref{EQ:Z0} is nonnegative.
This model is amenable to Monte Carlo simulation.
We now outline a simple and efficient algorithm for performing the stochastic sampling.

As a formal trick (in the spirit of Ref.~\onlinecite{Prokofev01}), 
we enlarge the phase space from $\Phi_0$ to 
$\Phi_0 \times \Phi_1 \times \cdots \times \Phi_N$,
where $\Phi_n$ is the set of configurations in which $2n$ free
endpoints have been introduced by breaking $n$ valence bonds.
(The system has been converted to one of both closed loops and open strings.)
We take the partition function to be
\begin{equation}
Z = \frac{1}{q^N} \sum_{C} q^{N_l(C)}\sigma^{N_s(C)} 
\prod_{[i,j] \in C} h_{ij}.
\end{equation}
The configurations $C$ are now assembled from all possible partial coverings
$v = ( [i_1,j_1], [i_2,j_2], \ldots, [i_n,j_n] )$ of variable length
$0 \le n \le N$, and $\sigma$ is introduced as a fugacity for the open strings [numbering $N_s(C) = N-n$].
The loop-only sector corresponds to the original partition function,
$Z_0 = \langle 1 \rangle_{\Phi_0}$.
(In each string sector there is a Green's function defined
by the string endpoints:
$G_{ij} = \langle \delta_{i,\alpha_1} \delta_{j,\beta_1} \rangle_{\Phi_1}$,
$G_{ij;kl} = \langle \delta_{i,\alpha_1} \delta_{j,\beta_1} \delta_{k,\alpha_2} 
\delta_{l,\beta_2} \rangle_{\Phi_2}$, etc.
Here, $\alpha_n$ and $\beta_n$ denote the positions of the head and tail
of the $n^{\text{th}}$ string. It is worth emphasizing that these $2n$-point
Green's functions do not coincide with expectation values of the physical
spin operators. In general, we must take all measurements 
in the $\Phi_0$ configuration space using the loop estimators 
derived in Ref.~\onlinecite{Beach06}.)

We will consider a process  that
involves breaking a single valence bond ($\Phi_0 \to \Phi_1$)
to produce an open string whose two endpoints (the ``head'' and ``tail'')
serve as walkers subject to Monte Carlo updates. The walkers
move via a series of two-step motions that involve drawing
a new bond and erasing an old one. When the walkers meet,
the loop is closed ($\Phi_1 \to \Phi_0$).
Figure~\ref{FIG:updates} shows an example circuit. 
The fives successive steps shown in panels (b)--(f)
produce an overall change in the relative weight
\begin{equation} \label{EQ:step5}
\frac{\sigma}{h_{5,4}} \times
\frac{h_{1,4}}{qh_{1,2}} \times
\frac{h_{5,8}}{qh_{7,8}} \times
\frac{qh_{3,2}}{h_{3,6}} \times
\frac{h_{7,6}}{\sigma}.
\end{equation}

Since we have chosen the bond amplitudes $h_{ij}$ to be nonnegative,
we can define a local amplitude $H_i = \sum_j h_{ij}$ and
a total overall amplitude $\mathsf{H} = \sum_{i} H_i = \sum_{ij} h_{ij}$.
These definitions will be useful in the derivations that follow.

To begin, let us consider processes that take the system from the space
of loops to the space of loops and one string. We move from a configuration
$C \sim [i,j]$ to a configuration $C' \sim (i)(j)$ by breaking a bond
$[i,j]$ and thus leaving string endpoints $(i)$ and $(j)$.
The transition probabilities for breaking and repairing the bond obey
the detailed balance equation
\begin{equation}
W^{\text{break}}_{[i,j]} P(i) \pi_C = W^{\text{repair}}_{(i)(j)}P(j | i) \pi_{C'}.
\end{equation}
Here $P(i)$ is the probability of choosing a site $i$ whose bond we want 
to break, and $P(j|i)$ is the probability of choosing $j$ given a walker
(string endpoint) at site $i$.
$\pi_C$ and $\pi_{C'}$ represent the likelihood of the system being found
in configurations $C$ and $C'$. Their ratio is given by
\begin{equation}
\frac{\pi_{C'}}{\pi_C} = \frac{\sigma}{h_{ij}}.
\end{equation}
If we choose which bond to break according to the distribution of local bond weight
$P(i) = H_i/\mathsf{H}$ and choose walker movements according to 
the distribution $P(j | i) = h_{ij} / H_i$, then
\begin{equation}
\delta = \frac{W^{\text{break}}_{[i,j]} }{W^{\text{repair}}_{(i)(j)}}   = \frac{P(j | i) \pi_{C'}}{P(i) \pi_C}
= \frac{\sigma}{\mathsf{H}}.
\end{equation}
We are free to choose $\sigma = \mathsf{H}$, in which case the transition probabilities
$W^{\text{break}}_{[i,j]}$ and $W^{\text{repair}}_{(i)(j)}$
are equal and unit-valued.

For motion of the walkers within $\Phi_1$,
we need to know the transition rates
between configurations $C\sim (i)[j,k]$ and $C' \sim [i,j](k)$. This represents
a process in which a walker at $i$ draws a new bond to some site $j$ and
then erases the preexisting bond connecting $j$ to $k$, thus leaving the
walker at site $k$. The detailed balance equation is
\begin{equation}
W^{\text{walk}}_{i\to k} P(j | i) \pi_C = W^{\text{walk}}_{k \to i} P(j | k) \pi_{C'}.
\end{equation}
The ratio
\begin{equation}
\frac{\pi_{C'}}{\pi_C} = q^{\delta N_l} \frac{h_{ij}}{h_{jk}}
\end{equation}
depends on $\delta N_l = N_l(C')-N_l(C) = \pm 1$ (or 0 if the moves do
not respect a fixed lattice bipartition; see discussion in Sect.~\ref{SECT:dynamic}).
As before, we attempt moves according to the
distribution $P(j | i) = h_{ij} / H_i$. Then,
\begin{equation}
\delta = \frac{ W^{\text{walk}}_{i\to k} }{ W^{\text{walk}}_{k\to i} } = \frac{P(j|k)}{ P(j|i) } \frac{\pi_{C'}}{\pi_C}
 = \frac{H_i}{H_k} q^{\pm 1},
\end{equation}
which can be solved in the usual way as $W^{\text{walk}}_{i\to k} = \delta/(1+\delta)$
or $W^{\text{walk}}_{i\to k} = \min(1,\delta)$.

Note that the transition rate does not depend on the ratio of bond amplitudes,
as it would if we had, for example, selected a site uniformly with $P(j | i) = 1/N$. 
The ratio $h_{ij}/h_{jk}$ may fluctuate wildly over many orders of magnitude,
so subsuming it into the sampling maximizes the efficiency of the algorithm.

In the case of a translationally invariant system, the amplitude for pairing spins at 
$i$ and $j$ must be a function of the vector $\mathbf{r}_{ij}$ connecting the two sites;
i.e., $h_{ij} = h(\mathbf{r}_{ij})$.
Hence, $H = \mathsf{H}/N = H_i = \sum_{\mathbf{r}} h(\mathbf{r})$  for all $i$, which implies that
$P(i) = H_i/\mathsf{H} \to 1/N$ is uniform and 
$P(j | i) = h_{ij}/H_{i} \to h(\mathbf{r}_{ij})/H$.
The algorithm can be summarized as follows:
\begin{enumerate}
\item Pick any valence bond $[i,j]$ (by choosing $i$ uniformly
from the set of A sublattice sites and then selecting its partner site in $v$ or $v'$)
and break it. The resulting string
has endpoints at $\mathbf{R} = \mathbf{r}_i$ and $\mathbf{R}' = \mathbf{r}_j$.

\item To move the head, choose a new bond vector $\mathbf{r}$ 
from the distribution $h(\mathbf{r})/H$.
So long as $\mathbf{R}+\mathbf{r} \neq \mathbf{R}'$,
attempt to draw a new bond from
$\mathbf{R}$ to $\mathbf{R}+\mathbf{r} = \mathbf{r}_k$ (for some $k$).
The bond $[k,l]$ that already exists at that site is then erased
and the walker is moved to $\mathbf{r}_l$.
The move is accepted with probability 1/2 if its effect is to join another
loop to the string and with probability 1 otherwise.

\item Otherwise, if $\mathbf{R} + \mathbf{r} = \mathbf{R}'$, close
the open string by drawing a new valence bond from $\mathbf{R}$ to $\mathbf{R}'$.
\end{enumerate}

\begin{figure*}
\ifthenelse {\boolean{PRBVERSION}}
{\includegraphics{dynamic.eps}}
{\includegraphics{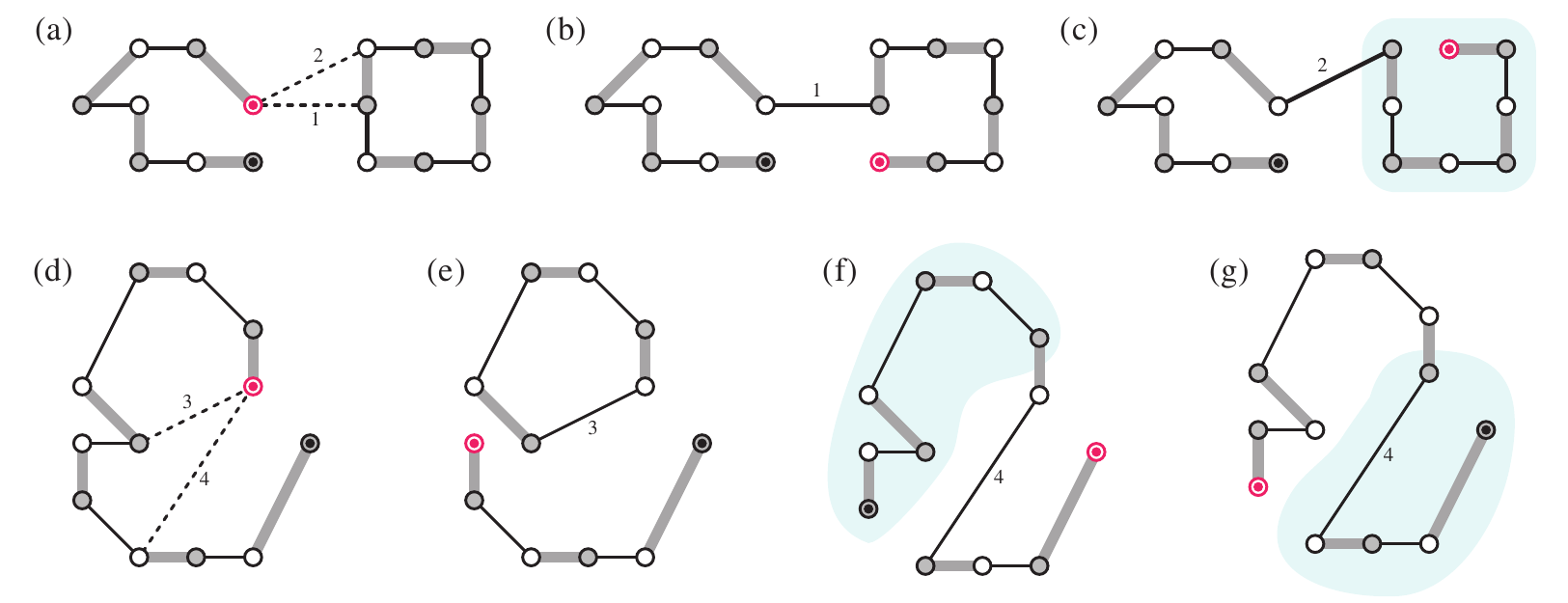}}
\caption{\label{FIG:dynamic}
\co
(a)~Two possible paths, marked 1 and 2, take the worm head to a site in the opposite or same sublattice
of another closed loop. In either case, the loop is absorbed.
(b)~Path~1 leads to a rearrangement of the worm that preserves the AB {\labelling}.
(c)~Path~2 requires that the AB {\labelling} be reversed in the highlighted region.
(d)~Another worm, following two possible paths marked 3 and 4.
(e)~For path~3, the AB {\labelling} is preserved, and the worm emits a new closed loop.
(f),(g)~Path~4 requires that the AB {\labelling} be reversed in the highlighted region. The number
of loops remains unchanged.
}
\end{figure*}

The worm algorithm described here is ergodic and guaranteed to have a high acceptance rate.
This is in contrast to the original bond-swap scheme proposed in Ref.~\onlinecite{Liang88},
wherein two A-site or B-site bond endpoints sitting diagonally across a plaquette
are swapped using Metropolis sampling. This antiquated algorithm runs into difficulty when the function
$h(\mathbf{r})$ is short ranged.
In particular, short bonds that are adjacent but not sharing a common plaquette generate long bonds under rearrangement, 
so whenever the amplitudes for long bonds become small, the acceptance rate can become correspondingly small.
Worse, there are typically many trapping configurations from 
which the simulation cannot emerge.
The worm algorithm does not suffer from these problems, because it can traverse any local barriers
by stepping outside the space of closed loops.
(We make no claims of novelty in this regard. Other approaches to overcome the sampling 
difficulty have been presented elsewhere.\cite{Albuquerque10,Tang11,Sandvik06})
  
\subsection{\label{SECT:dynamic}Fluctuating sublattice assignment}

The discussion in the previous section was specific to the case in which (i) the AB pattern is regular and (ii) the
$\mathbf{r}$ vectors that have nonzero $h(\mathbf{r})$ only connect sites in opposite sublattices. If those conditions hold, there
are only two possible consequences to the motion of the open string: a loop is joined to the string ($\delta N_l  = -1$) or  
a loop is split off from it ($\delta N_l = +1$). In both cases, represented
in Fig.~\ref{FIG:dynamic} by panels (a)$\to$(b) and (d)$\to$(e), the AB pattern itself is left undisturbed.

More generally, as the open string propagates it lays down a chain of singlet bonds whose alternating
site labels may be at odds with the traversed sites' current AB assignments. A simple workaround is to
flip the sublattice labels as required to correct the mismatch.
The relevant processes are now those in which 
a moving open string absorbs a closed loop 
($\delta N_l  = -1$) or reorganizes itself without impinging on any additional sites ($\delta N_l = 0$).
The first case is depicted in Fig.~\ref{FIG:dynamic} by panels (a)$\to$(c) and the second by 
(d)$\to$(f) or (d)$\to$(g). A crucial consideration is that, since the singlets are directional,
flipping sublattice labels along a loop segment has the effect of reversing a chain of singlet bonds.
If an odd number of singlets is effected, the overall sign of the wave function will change. 
This is true for all $\delta N_l = 0$ worm steps.

The sublattice mismatch can either be a temporary condition---lasting only until the worm updates succeed in laying
down a global AB pattern that is an invariant of the worm motion---or it may be that
the motion described by a given $h(\mathbf{r})$ is incompatible with any static AB site {\labelling}.
For example, consider the one-parameter family of short-range states on the square lattice described by
$h(\pm 1,0) = h(0,\pm 1) = \cos\theta$ and $h(\pm 1,\pm 1) = \sin\theta$ (with $0 \le \theta \le \pi/4$). Regardless of the initial 
sublattice {\labelling}---it can be any random assignment having an equal number of A and B labels---the
simulation will dynamically establish the checkerboard pattern provided that $\theta = 0$.
We keep track of the AB {\labelling} pattern by measuring a function 
$\Lambda(\mathbf{Q})=\sum_{\mathbf{r},\mathbf{r'}} e^{i\mathbf{Q}\cdot (\mathbf{r}-\mathbf{r'})} \langle\!\langle \lambda(\mathbf{r}) \lambda(\mathbf{r'}) \rangle\!\rangle$, where $\lambda(\mathbf{r})$ takes the value $-1$ or $1$ depending on the
current sublattice assignment at site $\mathbf{r}$. If $\theta = 0$, $\Lambda(\mathbf{Q})$ starts off broad
but systematically flows toward the distribution consisting of a single delta function peak at $\mathbf{Q} = (\pi,\pi)$; once
that is achieved, the pattern ceases to evolve.
Similar {\behaviour} is exhibited at $\theta = \pi/4$, where the system settles into a static pattern
with either $\mathbf{Q}=(\pi,0)$ or $\mathbf{Q}=(0,\pi)$. 
Only in those two extreme cases is the sublattice pattern eventually static and the simulation sign-problem free.

\section{\label{SECT:results}Results}

\begin{table*}[h!]
 \begin{tabular}{|r|r|r|r|r|r|r|r|} \hline
n  & $Z$ & $-L^2\,C_1$ & $L^2\,C_2$ & $L^4\,M^2(\pi,\pi)$ & $L^4\,M^2(\pi,0)$ & $L^2\,\lvert D \rvert$ & $L^4\, 4D^2 / 3$ \\ \hline\hline
0 & 1559232 & 22241280 & 9902080 & 113983488 & 17383424 & 4376064 & 102133760 \\
1 & 13008384 & 194568192 & 104726528 & 1117618176 & 139902976 & 28540928 & 645455872 \\
2 & 66018816 & 997232640 & 585695232 & 6104383488 & 709410816 & 127591424 & 2844606464 \\
3 & 223842816 & 3395051520 & 2137292800 & 21861335040 & 2381496320 & 389861376 & 8395677696 \\
4 & 568694016 & 8564477952 & 5689352192 & 57653526528 & 6069354496 & 932687872 & 19758309376 \\
5 & 1108661760 & 16547069952 & 11594661888 & 116342292480 & 11792498688 & 1697314816 & 35459866624 \\
6 & 1767412224 & 25797685248 & 18932629504 & 189239033856 & 18888998912 & 2580870144 & 53692563456 \\
7 & 2302253568 & 32679444480 & 25148850176 & 250229981184 & 24519589888 & 3165620224 & 65523884032 \\
8 & 2528419968 & 34418749440 & 27661209600 & 275349995520 & 27030159360 & 3329164288 & 68794482688 \\
9 & 2302253568 & 29878050816 & 25148850176 & 250229981184 & 24519589888 & 2857185280 & 58976903168 \\
10 & 1767412224 & 21512073216 & 18932629504 & 189239033856 & 18888998912 & 2089987072 & 43192369152 \\
11 & 1108661760 & 12538503168 & 11594661888 & 116342292480 & 11792498688 & 1223688192 & 25387999232 \\
12 & 568694016 & 5848903680 & 5689352192 & 57653526528 & 6069354496 & 594391040 & 12360392704 \\
13 & 223842816 & 2070282240 & 2137292800 & 21861335040 & 2381496320 & 218601472 & 4580990976 \\
14 & 66018816 & 528863232 & 585695232 & 6104383488 & 709410816 & 60980224 & 1313734656 \\
15 & 13008384 & 84836352 & 104726528 & 1117618176 & 139902976 & 11331584 & 254992384 \\
16 & 1559232 & 6254592 & 9902080 & 113983488 & 17383424 & 1074688 & 31887360 \\ \hline
\end{tabular}
\caption{\label{TAB:coefficients} The integer coefficients appearing as $z_n$ and $o_n$ in Eq.~\eqref{EQ:rational_polynomial}
are presented for various observables. These coefficients specify the rational polynomials in $x=h(2,1)/h(1,0)$ that 
arise from taking expectation values with respect to the product amplitude trial state on the square lattice
of linear size $L=4$.
The columns correspond to the wave function normalization, the nearest- and next-nearest-{\neighbour} 
spin correlations, the staggered and stripe magnetization, and the columnar dimer order
parameter (with measurements of both its absolute value and its square).
}
\end{table*}

\begin{figure}
\begin{center}
\ifthenelse {\boolean{PRBVERSION}}
{\includegraphics{factoradic.eps}}
{\includegraphics{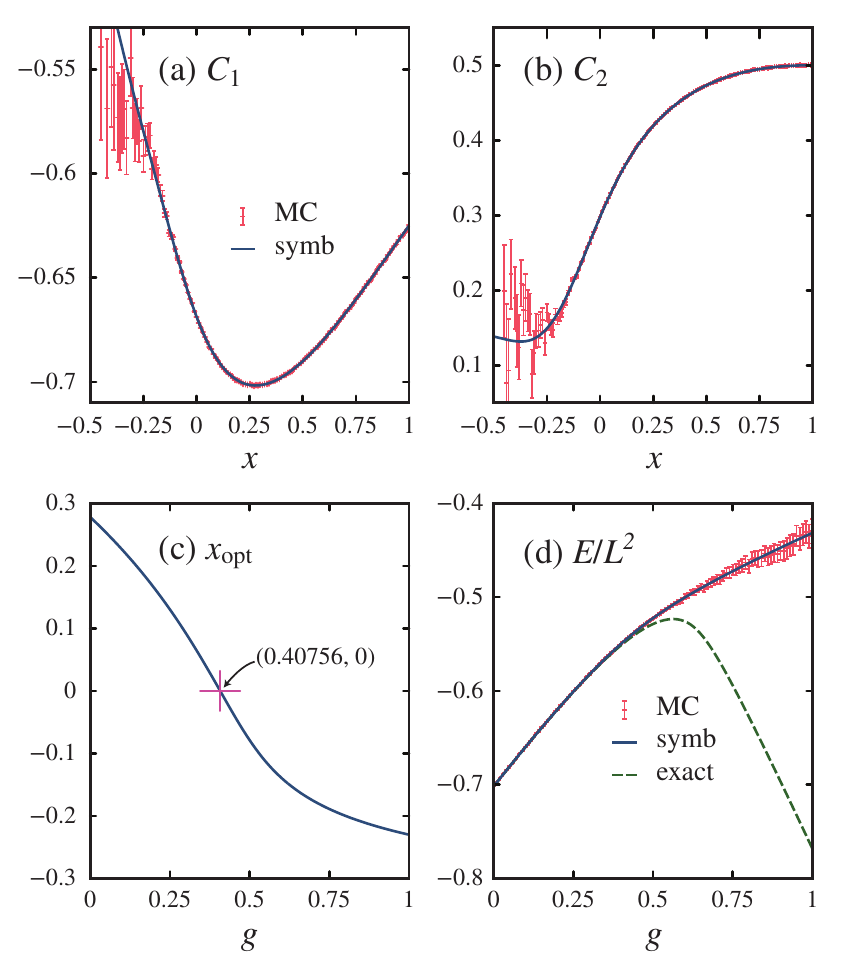}}
\end{center}
\caption{\label{FIG:factoradic} \co 
RVB trial wave function results for the $4\times 4$ lattice.
(a),(b)~Spin correlations $C_1$ and $C_2$ between nearest- and next-nearest-{\neighbour} spins, computed 
as a function of the amplitude ratio $x=h(2,1)/h(1,0)$. 
The worm Monte Carlo (MC) results are compared to the corresponding symbolic (symb) expression.
As $x$ becomes increasingly negative, the stochastic evaluation becomes dominated by noise from the sign problem.
(c)~The energy-optimized value $x_{\text{opt}}$ remains
positive up to $g=J_2/J_1 = 0.40756$. 
(d)~The optimized trial state gives a good approximation to the true ground state
energy (exact) up to where the Marshall sign rule breaks down.
}
\end{figure}

As a test of the worm implementation, we compare its output to \emph{analytical} results obtained
for the $4\times 4$ lattice. We exploit the fact that the bipartite valence bond basis $\mathcal{V}_{\text{AB}}$
for $2N$ spins is isomorphic to the set of permutations on $N$ elements.\cite{Beach06}
Hence, the basis states have a natural lexical ordering via the Lehmer code\cite{Lehmer60,Knuth73} and can easily be enumerated.
For $4\times 4 = 16$ sites, the total number of the states is only $8! = 40\,320$, which means that expectation
values of the trial wave function can be evaluated exactly at very little
computational cost.
Moreover, we can carry out the calculation symbolically. Each observable takes the form of
a rational function of order [16/16]:
\begin{equation} \label{EQ:rational_polynomial}
\langle \hat{O} \rangle = \frac{O(x)}{Z(x)} = \frac{3}{4} \frac{\sum_{k=0}^{16} o_k x^k}{\sum_{l=0}^{16} z_l x^l}.
\end{equation}
The argument of the polynomials appearing in the numerator and denominator is the real-valued ratio $x = h(2,1)/h(1,0)$, and the coefficients $o_k$ and $z_k$ are all integers.
Specific values for various observables are listed in Table~\ref{TAB:coefficients}.

For this test we have {\focussed} on the nearest- and next-nearest-{\neighbour} spin correlation functions,
$C_1=\frac{1}{L^2}\sum_{\langle i,j\rangle}\langle\mathbf{S}_i\cdot \mathbf{S}_{j}\rangle $ and
$C_2=\frac{1}{L^2}\sum_{\langle \langle i,j\rangle \rangle} \langle \mathbf{S}_i\cdot \mathbf{S}_{j} \rangle$;
the $\mathbf{Q} = (\pi,\pi)$ staggered and $\mathbf{Q} = (\pi,0)$ stripe magnetization, 
$M^{2}({\mathbf{Q}})=\frac{1}{L^4}\sum_{\mathbf{r,r'}}(-1)e^{i\mathbf{Q}\cdot(\mathbf{r}-\mathbf{r}')}\langle \mathbf{S}_{\mathbf{r}}\cdot \mathbf{S}_{\mathbf{r}'}\rangle$; and the order
parameter for a columnar dimer crystal,
$D^2=\frac{1}{L^4}\sum_{\mathbf{r,r'}}(-1)^{\mathbf{e}_x\cdot(\mathbf{r}+\mathbf{r}')}\langle (\mathbf{S}_{\mathbf{r}}\cdot
\mathbf{S}_{\mathbf{r}+\mathbf{e}_x})(\mathbf{S}_{\mathbf{r}'}\cdot \mathbf{S}_{\mathbf{r}'+\mathbf{e}_x})\rangle$.
We have verified that the worm algorithm, conventional bond swap Monte Carlo, and exact evaluation 
give consistent results for all these quantities. 

The comparison of the energetics is shown in Fig.~\ref{FIG:factoradic}. 
Note that in Figs.~\ref{FIG:factoradic}(a) and \ref{FIG:factoradic}(b), 
the stochastic evaluation of $C_1$ and $C_2$ continues to work in some range of $x < 0$ but breaks down as $x$ becomes
strongly negative. For the symbolic result, the determination of the best energy is carried out by considering
the two-parameter function $\mathcal{E}(x,g)/J_1L^2 = C_1(x) + gC_2(x)$, 
which is known exactly by way of Eq.~\eqref{EQ:rational_polynomial}.
For every value of the relative coupling strength $g$, the optimal value of $x$ [Fig.~\ref{FIG:factoradic}(c)]
is the one that produces the lowest energy [Fig.~\ref{FIG:factoradic}(d)] according to
\begin{equation} \label{EQ:energy_min_4x4}
E(g) = \mathcal{E}(x_{\text{opt}},g) = \underset{x}{\text{min}}\, \mathcal{E}(x,g).
\end{equation}
In practice, Eq.~\eqref{EQ:energy_min_4x4} represents
a root-finding problem in $x$ for $\partial \mathcal{E}(x,g)/\partial x = 0$; this is
solved via Newton-Raphson. We find that the optimized value $x_{\text{opt}}$ is positive for weak frustration.
It decreases monotonically from its nonfrustrated value, $x_{\text{opt}} = 0.2780138519$, and drops below zero when the coupling 
strength exceeds $g = 0.40756$. This marks the point at which the Marshall sign rule first fails. For reference (it may be
of use in benchmarking RVB calculations accomplished by other methods, e.g., Ref.~\onlinecite{Wang13}), 
we report that the specific values $x_{\text{opt}} = 0.006787458952$, $-0.03777121711$, $-0.07881072679$, and $-0.1128184711$
obtain at coupling strengths $g = 0.40$, $0.45$, $0.50$, and $0.55$.

\begin{figure}
\begin{center}
\ifthenelse {\boolean{PRBVERSION}}
{\includegraphics{crossings.eps}}
{\includegraphics{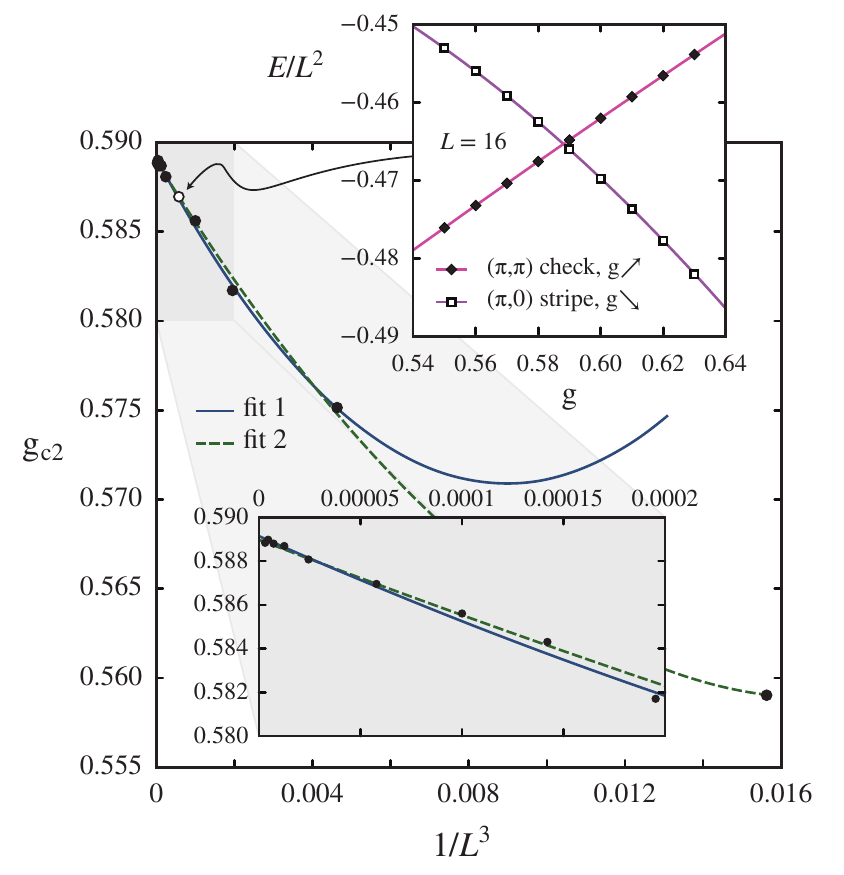}}
\end{center}
\caption{\label{FIG:crossings}\co The level crossings are plotted versus $1/L^3$ 
and extrapolated to the thermodynamic limit. Several different 
second-order polynomial fits (two shown) are used to estimate the uncertainty in the intercept. 
The solid, blue line (fit 1) is an attempt to fit the $L \ge 6$ data to 
$c_0\exp(c_1L^{-3} + c_2L^{-6})$;
the dashed, green line (fit 2) is a fit to $c_0 + c_1L^{-3} + c_2L^{-6}$ for $L \ge 4$.
Our analysis suggests a value $g_{\text{c}2} \doteq 0.5891(3)$. The upper inset shows the analysis
behind the $L=16$ data point, which is marked in the main graph as an open circle. The lower inset is
a magnification of the shaded region.
}
\end{figure}

\begin{figure}
\begin{center}
\ifthenelse {\boolean{PRBVERSION}}
{\includegraphics{phases.eps}}
{\includegraphics{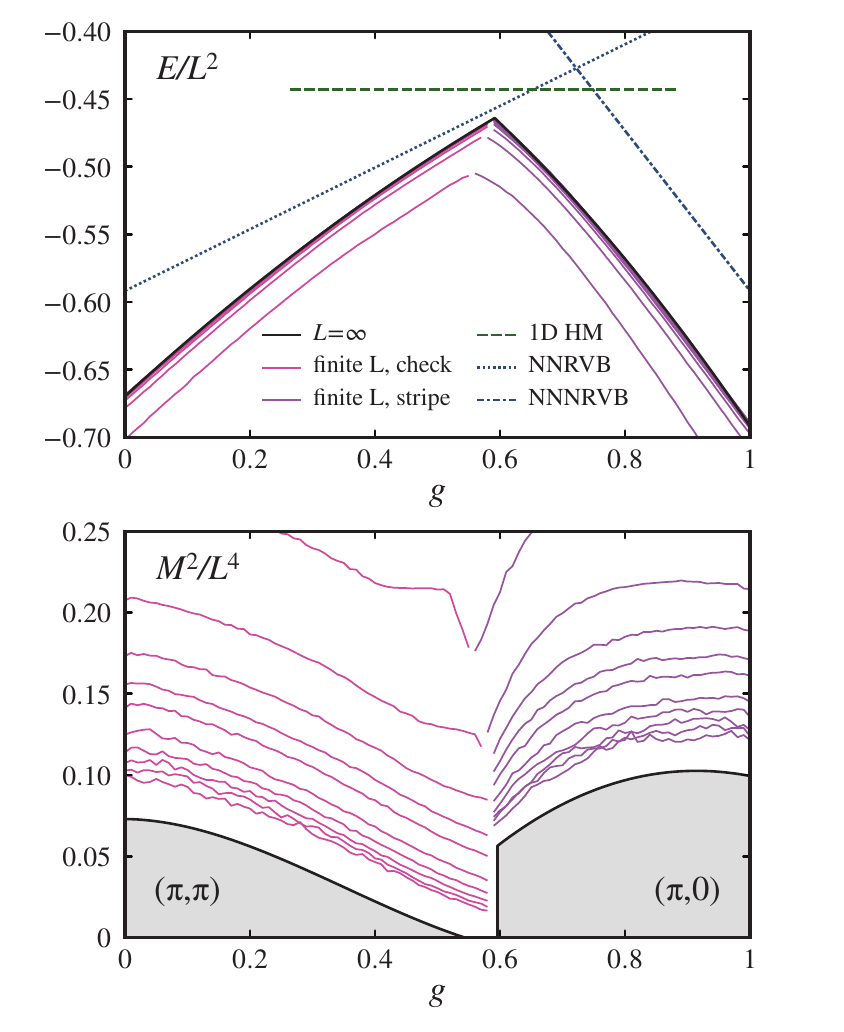}}
\end{center}
\caption{\label{FIG:phases} \co
(Upper panel)~The energy per site versus the coupling strength. The solid
lines are best energies from the trial wave function optimization.
From bottom to top, system sizes $L = 4, 6, 8, 10, 12, 16, 20, 24, 28, 32$
are shown. The thick black line, providing an upper envelope to the curves, is the
extrapolation to $L=\infty$. The ground state energy of the one-dimensional Heisenberg chain is shown for comparison,
as are the energies of the NNRVB state and its $45^{\circ}$-rotated, next-nearest-{\neighbour} {\analogue}, the NNNRVB state.
(Lower panel)~Magnetization data are shown, with the same system sizes now increasing
from top to bottom. The thick black lines above the {\grey} shading are the $L=\infty$
extrapolation. The magnetic order for $\mathbf{Q} = (\pi,\pi)$ and $\mathbf{Q} = (\pi,0)$ both
vanish in the small region between $g_{\text{c}1} \doteq 0.54(1)$
and $g_{\text{c}2} \doteq 0.5891(3)$.
}
\end{figure}

\begin{figure}[b]
\begin{center}
\ifthenelse {\boolean{PRBVERSION}}
{\includegraphics{marshall.eps}}
{\includegraphics{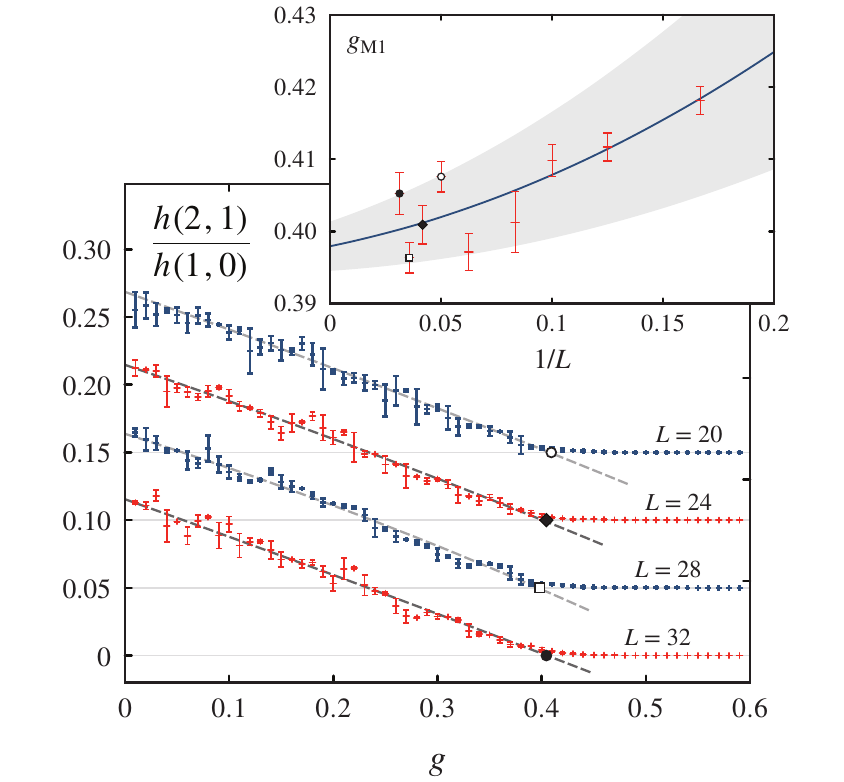}}
\end{center}
\caption{\label{FIG:marshall} \co (Main panel) The knight's move amplitude measured relative to the nearest-{\neighbour}
bond amplitude---offset vertically by $0.05 \times (8-L/4)$ to aid viewing---decreases as a function of $g$.
(Inset) The coupling strength at which $h(2,1)$ extrapolates to zero is plotted against the inverse linear 
system size. The point style for each system size matches the intercept in the main panel. 
The shaded region represents the envelope containing plausible fits. We estimate that
the checkerboard Marshall sign rule fails at $g_{\text{M}1} \doteq 0.398(4)$ in the thermodynamic limit.
}
\end{figure}

\begin{figure}
\begin{center}
\ifthenelse {\boolean{PRBVERSION}}
{\includegraphics{amplitudes.eps}}
{\includegraphics{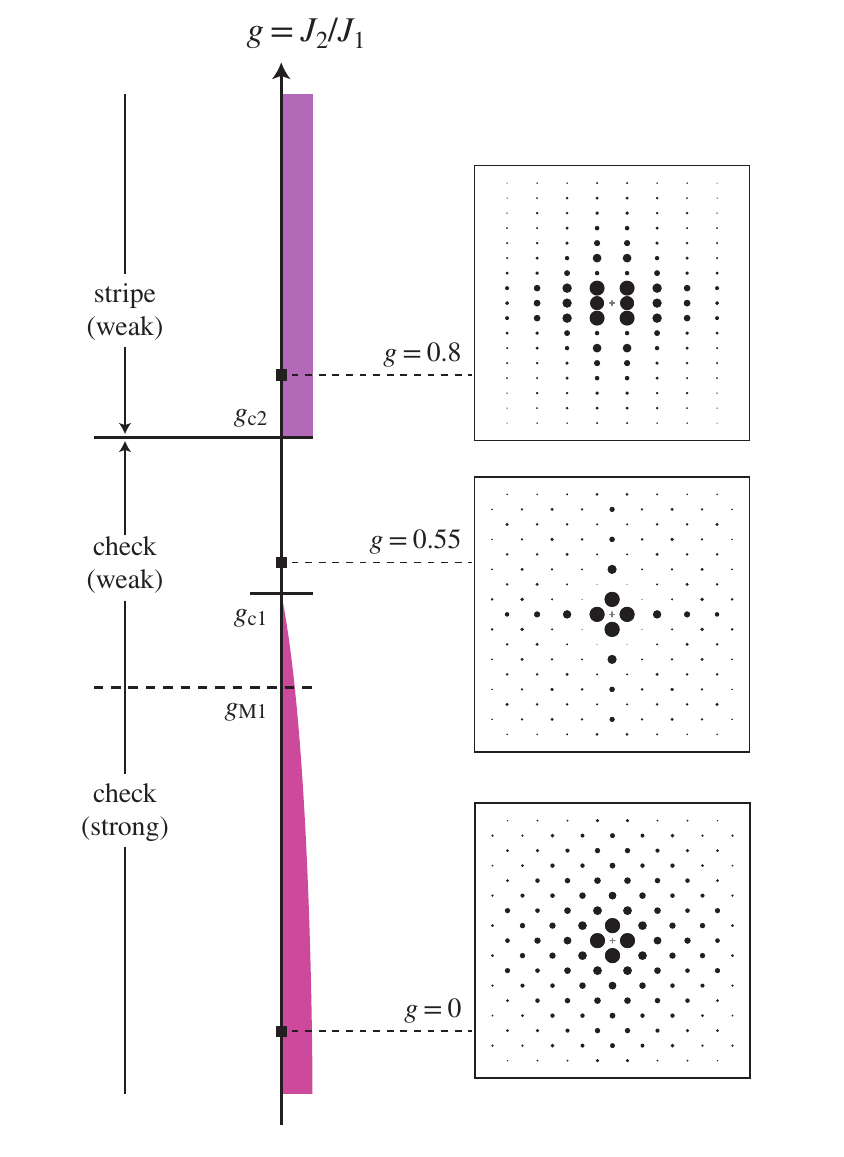}}
\end{center}
\caption{\label{FIG:amplitudes} \co Schematic representation of
the model's zero temperature phase diagram. Critical couplings $g_{\text{c}1}$
and $g_{\text{c}2}$ mark the boundaries of the magnetically disordered phase.
Staggered order ends with a continuous transition at $g_{\text{c}1}$; stripe order
ends with a first-order transition at $g_{\text{c}2}$. The three diagrams on the
right illustrate the optimized $h(\mathbf{r})$ values at $g=0$, $g=0.55$, and $g=0.8$.
Each circle, offset by a vector $\mathbf{r}$ (measured from the small cross at the {\centre}), has an area proportional
to the corresponding $h(\mathbf{r})$ value. The text on the left describes the
Marshall sign structure that predominates.
}
\end{figure}

Having established confidence in our numerical implementation, we proceed with
unbiased optimization calculations using a static sublattice assignment on lattices up
to size $L=32$. Convergence is limited by statistical uncertainty in the (energy to bond count)
correlation function that determines the local energy gradient,\cite{Lou07}
and it is difficult to optimize reliably for larger system sizes.
(See Appendix~\ref{SEC:optimization} for more details.)
We first consider the checkerboard AB pattern. 
At $g=0$, the bond amplitudes are given an initial value
\begin{equation}
h(x,y) = \bigl[\min(x,L-x)^2 + \min(y,L-y)^2\bigr]^{-3/2}
\end{equation}
for $\lvert x \rvert + \lvert y \rvert$ odd and zero otherwise.
The new set of amplitudes obtained from this first run serves as the input for the next optimization process. 
That is to say, we daisy chain the calculations, at each step using the converged result at $g$ to seed the simulation at $g + \delta g$.
An analogous procedure is carried out for the stripe AB pattern, starting from $g=\infty$ and stepping the relative
coupling down.

One finds that the two sets of simulations do not join smoothly but instead meet with strongly opposite
slopes $dE/dg$. A careful extrapolation to the thermodynamic limit,
presented in Fig.~\ref{FIG:crossings}, puts the location of the energy level crossing at $g_{\text{c}2} \doteq 0.5891(3)$. 
As Fig.~\ref{FIG:phases} makes clear, this point represents the rightmost edge of an intermediate phase that
is magnetically disordered. The leftmost edge sits at $g_{\text{c}1} \doteq 0.54(1)$, where the $\mathbf{Q}=(\pi,\pi)$
antiferromagnetism vanishes in a continuous fashion. As a rough gauge of the quality of the RVB
trial wave function, we note that for $g=0.5$ the energy density extrapolates 
 to $E_{\text{RVB}} = -0.49023(2)$ in the thermodynamic limit.
This result is bracketed by the energies of the best projected entangled pair states (PEPS) 
with bond dimension $D=3$ [$E_{\text{PEPS}} = -0.48612(2)$; see Ref.~\onlinecite{Wang13}] and $D=9$
[$E_{\text{PEPS}} = -0.4943(7)$; see Ref.~\onlinecite{Wang11}].

An important detail is that the optimizations are carried out with the bond amplitudes constrained to have $x$- and 
$y$-axis reflection symmetry but not necessarily $90^{\circ}$ rotation symmetry. In the case of the checkerboard simulation, 
the amplitudes nonetheless realize the full lattice symmetry under optimization up to large values of
the relative coupling. For small lattice sizes $L=4, 6, 8$, the symmetry 
breaks down beyond values $g \approx 0.51, 0.55, 0.57$. For all larger sizes, that point is pushed well to the right 
of $g_{\text{c}2}$. This means that, in the thermodynamic limit, 
$h(\mathbf{r})$ shares a common symmetry across both the staggered magnetic
phase and the disordered intermediate phase. But it experiences a sudden break at the onset of stripe
magnetic order, dropping from $C_4$ to $C_2$.

In the vicinity of $g=0$, the optimized bond amplitudes are positive definite and an almost perfect function of 
bond length. As the frustration increases, the amplitudes begin to deviate from circular symmetry, developing
strong lobes of weight along the x and y axes. Bonds not aligned along those preferred directions become increasingly short
ranged, and the eight knight's move bonds, those symmetry equivalent to $h(2,1)$, eventually 
trend through zero to negative values.
The extrapolation shown in Fig.~\ref{FIG:marshall} pinpoints the breakdown of the Marshall sign rule at
$g_{\text{M}1} \doteq 0.398(4)$. What this suggests is that there is strict adherence to a checkerboard
Marshall sign rule only below $g_{\text{M}1}$; in the range $g_{\text{M}1} < g < g_{\text{c}2}$, the sign rule
is violated, even though the overall sign structure is still partially consistent with the checkerboard pattern. 
[There is no indication that the amplitudes of any other bond type are on track to change sign. Attempts to
extrapolate the amplitudes next most likely to turn negative, viz.\ $h(4,1)$ and $h(6,1)$, 
put their vanishing points deep in the intermediate phase or beyond it.]
We find that the {\behaviour} on the large coupling side is not comparable. There, the coupling at which bond amplitudes
first go negative scales as $g_{\text{M}2} \sim L^4 $ and hence does not converge in the thermodynamic limit. 
We interpret this to mean that the static stripe pattern is only ever a weak description of the Marshall sign structure. 
See Fig.~\ref{FIG:amplitudes}.

We have attempted to confirm this picture by running simulations in which the Marshall sign structure is determined
dynamically. More specifically, we want to verify that the strongly first-order transition at $g_{\text{c}2}$ is not merely
an artifact of two static, incompatible sublattice conventions colliding. And we would like to see if any pattern
other than checkerboard or stripe could emerge on its own. If permitted, might the system's sublattice structure
smoothly interpolate over some range of $g$, with the peak in $\Lambda(\mathbf{Q})$ migrating from $(\pi,\pi)$ to $(\pi,0)$?
Or perhaps with the peak in $\Lambda(\mathbf{Q})$ broadening into incoherence?
We follow the procedure outlined in Sect.~\ref{SECT:dynamic}, whereby the sublattice {\labelling}
is no longer fixed and the worm motion itself is allowed to reconfigure the current AB pattern. 
Our approach is to simulate for various $g$ values---with \emph{no} daisy chaining---in each case
starting from a random AB pattern and a random loop configuration.
The bond amplitudes are initialized with $h(\mathbf{r})$ forming a broad peak around $\mathbf{r} = \mathbf{0}$ and having no zero
entries. We perform a crude simulation in which the signs associated with the worm updates
are thrown away. (See Appendix~\ref{SEC:sign_problem}.) Otherwise, the optimization of $h(\mathbf{r})$ proceeds as before.
What we find is a result that exactly tracks the state of lower energy produced by assuming one of the two static AB patterns.
The simulation flows to the checkerboard for all $g < g_{\text{c}2}$ and to the stripe for all $g >g_{\text{c}2}$; the peak
in $\Lambda(\mathbf{Q})$ jumps discontinuously. Obviously we should not read too much into a result
that follows from an uncontrolled approximation (sampling by ignoring the signs), but it does give us a sense that the 
stability of the checkerboard pattern through the intermediate phase and the 
abrupt change in Marshall sign structure at $g_{\text{c}2}$ might be genuine features of the model.

\begin{figure}
\ifthenelse {\boolean{PRBVERSION}}
{\includegraphics{dimer_pattern.eps}}
{\includegraphics{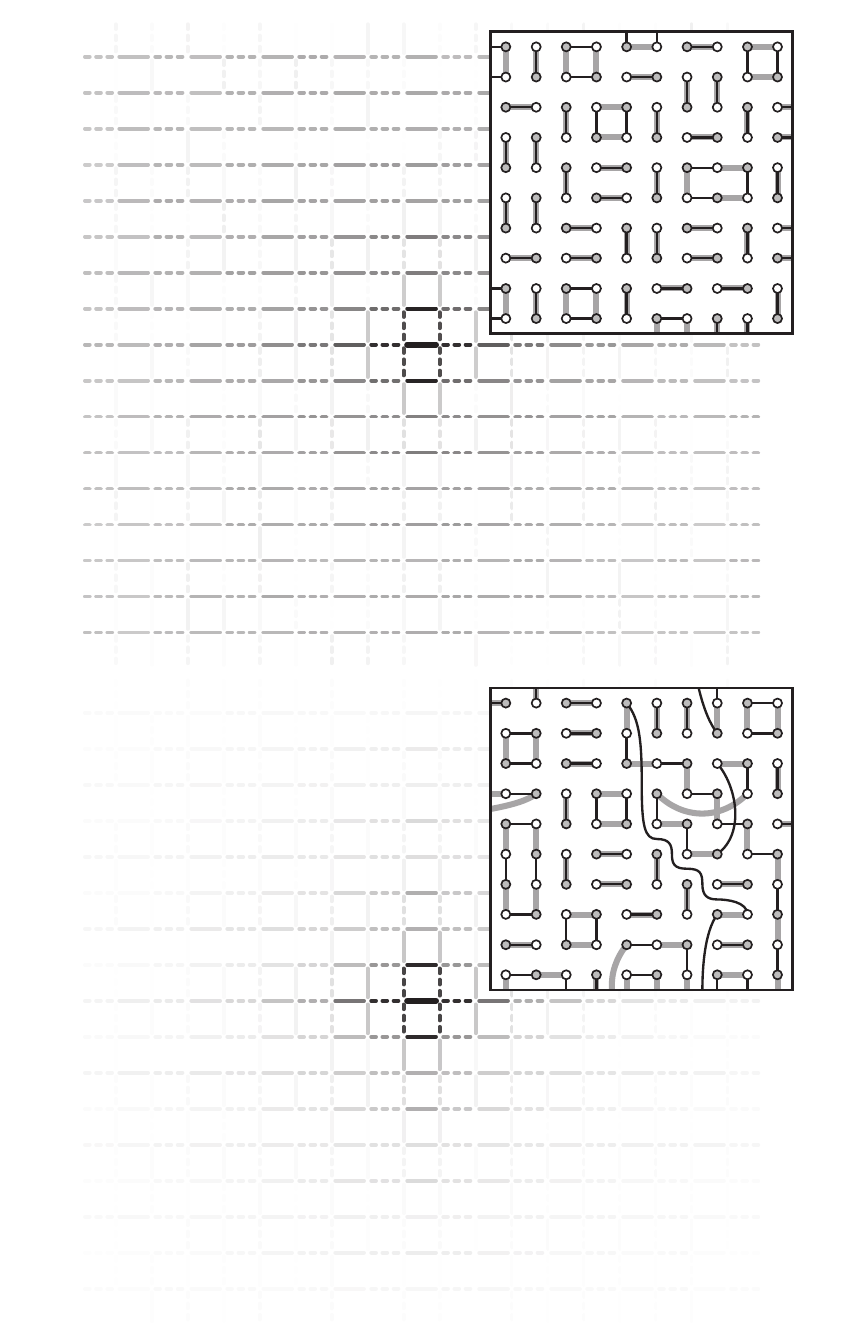}}
\caption{\label{FIG:dimer_pattern}
Grid lines depict the dimer correlations $C_{ijkl} = \langle (\mathbf{S}_i\cdot\mathbf{S}_j)(\mathbf{S}_k\cdot\mathbf{S}_l)\rangle$
on the nearest-{\neighbour} links $(k,l)$ of the square lattice, measured with respect to the thick, dark dimer $(i,j)$ at the {\centre}. 
The correlations are computed for the $L=28$ system.
The {\greyscale} intensity represents correlation
strength---presented as the fourth power of $(1+\tfrac{3}{2}r_{ij;kl}^{3/2})C_{ijkl}$, where 
$r_{ij;kl}$ is the distance measured from the {\centre} of the $(i,j)$ bond to the {\centre} of the
$(k,l)$ bond. Dotted lines indicate a negative (anticorrelated) value. The top panel shows results for the NNRVB state, 
presented for comparison's sake. The bottom panel shows results for the energy-optimized state at $g=0.58$.
In each case, a $10\times 10$ section of the full valence bond loop configuration,
obtained from a snapshot of the Monte Carlo simulation, is overlaid. 
}
\end{figure}

\begin{figure}
\ifthenelse {\boolean{PRBVERSION}}
{\includegraphics{dimer_corr.eps}}
{\includegraphics{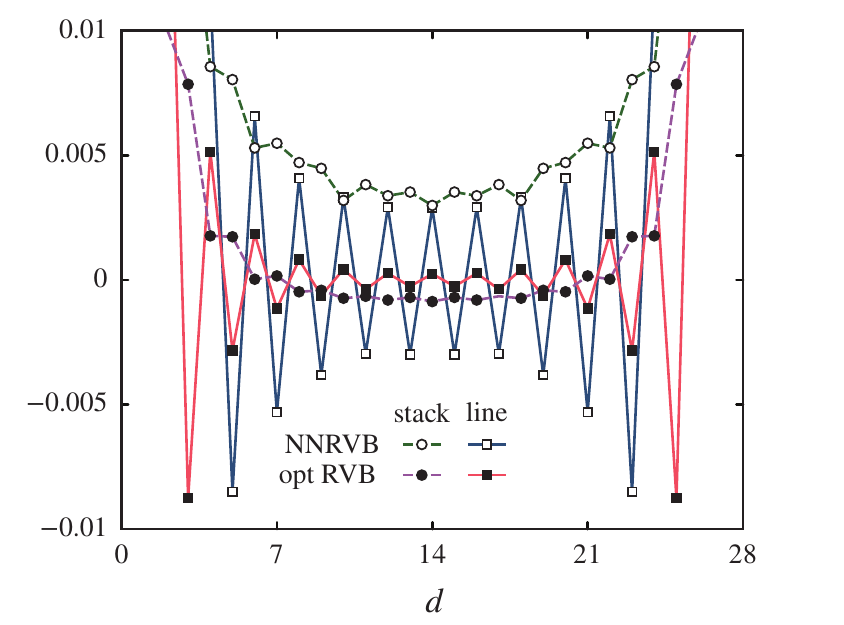}}
\caption{\label{FIG:dimer_corr} \co Dimer correlations of the product-amplitude trial
wave function optimized at $g = 0.58$ (solid points) and the short-bond-only RVB state (open points)
are compared on the $L=28$ lattice. Presented are the dimer line (squares)
and dimer stack (circles) correlation functions. See Eq.~\eqref{EQ:dimer_correlation_functions} 
and accompanying text for definitions.
}
\end{figure}

The optimized state in the intermediate phase is definitely not a bond crystal. For a given lattice, 
the dimer correlations are somewhat enhanced in the strongly frustrated region, but with increasing lattice size 
they show clear convergence to zero. Still, spatially resolved dimer correlations do give us important
information. One can see in Fig.~\ref{FIG:dimer_pattern} that the optimized state shows the
same pattern of dimer correlation and anticorrelation as the NNRVB, but it decays
much faster as a function of dimer separation. The comparison is made more explicit in Fig.~\ref{FIG:dimer_corr},
which shows correlations along a line and a stack of dimers.
The functions measured are
\begin{equation} \label{EQ:dimer_correlation_functions}
\begin{split}
C_\text{line}(d) &= \langle \hat{B}(0,0)\hat{B}(d,0) \rangle - \langle \hat{B}(0,0)\rangle\langle \hat{B}(d,0) \rangle, \\
C_\text{stack}(d) &= \langle \hat{B}(0,0)\hat{B}(0,d) \rangle - \langle \hat{B}(0,0)\rangle\langle \hat{B}(0,d) \rangle,
\end{split}
\end{equation}
which we have expressed in terms of the $x$-directed bond operator $\hat{B}(x,y) = \mathbf{S}(x,y)\cdot\mathbf{S}(x+1,y)$.

\section{\label{SEC:conclusions} Conclusions}

We have used an optimized valence bond  trial wave function to study the square-lattice $J_1$--$J_2$ Heisenberg model, 
with an eye to both mapping out the zero-temperature
phase diagram and determining how the Marshall sign structure breaks down near the phase boundaries.
In the first instance, we fix the AB sublattice {\labelling} to coincide with
the  order that exists at small and large coupling. For each lattice size, the 
intermediate phase is approached in two independent simulations (or, rather, chains of history-dependent simulations)
by evolving the states progressively out of the two ordered phases, minimizing their energy at each step.
These simulations are fully non-sign-problematic, since the AB pattern is fixed and the bond amplitudes
are restricted to be positive.

Finite-size scaling of the dimer order parameter suggests that there is no long-range dimer order at any value of $g$.
This is as expected, since the trial state explicitly ignores bond-bond correlations beyond those generated
by the hardcore tiling constraint. Measurements of the staggered magnetization show clear evidence of a continuous phase transition 
in which the staggered magnetization vanishes
at $g_{\text{c}1} \doteq 0.54(1)$. On the other edge of the intermediate phase, an energy level crossing
at $g_{\text{c}2} \doteq 0.5891(3)$ results in the sudden disappearance of the otherwise robust stripe magnetization.
This is accompanied by the restoration of the system's rotational symmetry.
(Since the trial state is least able to describe the intermediate phase---again, because of its lack of explicit
bond-bond correlations---we should probably view $g_{\text{c}1}$ and $g_{\text{c}2}$ as upper and lower
bounds, respectively, on the true positions of the phase boundaries.)
We have also performed calculations (approximate and uncontrolled, but suggestive) in 
which no sublattice {\labelling} is put in by hand and the AB pattern is allowed to emerge dynamically.
We find that, regardless of the initial sublattice assignment, the simulation reliably settles into the checkerboard 
pattern for all $g < g_{\text{c}2}$ and the stripe for all $g > g_{\text{c}2}$.
Taken together, our results point to the checkerboard AB pattern being the best choice 
throughout the intermediate phase. Hence, within the context
of our particular trial wave function scheme, we surmise that the state beyond $g_{\text{c}1}$ is a ``bosonic''  spin liquid
with the lowest-lying magnetic excitations at $(\pi,\pi)$. 

Figure~\ref{FIG:amplitudes} gives a quick summary of our results. We observe that at high frustration the bond amplitudes take
on a highly anisotropic form. This is quite different from the long-bond to short-bond picture that is usually invoked. 
Recall that Liang, Ducot, and Anderson studied long-range RVB states on the square lattice with 
amplitudes $h \sim r^{-p}$ that decay as a power law in the bond length $r$.~\cite{Liang88} 
In that framework, the state becomes magnetically disordered when
$p$ exceeds a critical value of 3.3,~\cite{Havilio99,Havilio00,Beach07b} 
and the entire family of states in the range $p > 3.3$ is continuously connected to $p=\infty$,
which is the (short-bond-only) NNRVB. The intermediate phase state obtained in our simulations is of
a quite different character: (i) the state is magnetically disordered not because its bond amplitudes are uniformly short 
ranged but because they have become short ranged over some sufficiently large angular interval of bond orientation;
(ii) its spin and dimer correlations are distinct from those of the NNRVB;
and (iii) the presence of many system-spanning bonds implies that the usual topological
invariant for short-ranged RVB states, defined by the parity of bond cuts along a reference line,\cite{Diep05,Albuquerque10}
is almost certainly not a good quantum number.

This work was supported by a Discovery grant from NSERC of Canada.

\appendix

\section{\label{SEC:optimization}Numerical optimization of the RVB bond amplitudes}

The RVB ansatz assumes that the quantum amplitude $\psi(v)$ associated with each valence bond state $| v \rangle$
is of the factorizable form
\begin{equation} \label{EQ:rvb_ansatz}
\psi(v) \approx \prod_{[i,j] \in v} h(\mathbf{r}_{ij}) \equiv \prod_{\tilde{\mathbf{r}}} h(\tilde{\mathbf{r}})^{n(\tilde{\mathbf{r}};v)}.
\end{equation}
The first product ranges over all pairs of spins forming a singlet bond. The second 
ranges over the minimal set of vectors $\tilde{\mathbf{r}}$ that are inequivalent under whatever lattice symmetries 
have been enforced. The whole-number exponent $n(\tilde{\mathbf{r}};v)$ represents
how many times a bond amplitude $h(\mathbf{r})$, with $\mathbf{r}$ symmetry-equivalent to $\tilde{\mathbf{r}}$,
appears in the product for a given $v$.
[Hence, $\sum_{\tilde{\mathbf{r}}} n(\tilde{\mathbf{r}}) = L^2/2 = N$, the number
of bonds appearing in $| v \rangle$.]

Accordingly, the energy expectation value is
\begin{equation} \label{EQ:energy_expectation_value}
E = \frac{\langle \psi | \hat{H} | \psi \rangle}{\langle \psi |  \psi \rangle} 
= \frac{\sum_{C} H(C) w(C) }{\sum_{C} w(C)} \equiv \langle\!\langle H \rangle\!\rangle,
\end{equation}
where $H(C) = \langle v | \hat{H} | v' \rangle /  \langle v | v' \rangle$
is the loop estimator of the Hamiltonian. The 
notation $\langle\!\langle \cdot  \rangle\!\rangle$ denotes averaging with respect to the Monte Carlo weight
\begin{equation} \label{EQ:Monte_Carlo_weight}
w(C) =  \langle v | v' \rangle\psi(v)\psi(v')  = \pm q^{N_l(C)}\prod_{\tilde{\mathbf{r}}}   h(\tilde{\mathbf{r}})^{n(\tilde{\mathbf{r}};C)}.
\end{equation}
Here, each configuration $C = (v,v')$ is a superposition of two dimer coverings,
and the sum $n(\tilde{\mathbf{r}};C) \equiv n(\tilde{\mathbf{r}};v)+n(\tilde{\mathbf{r}};v')$ is the combined count  of 
$\tilde{\mathbf{r}}$-type bonds in states $| v \rangle$ and $| v' \rangle$.
The $\pm$ on the right-hand-side of Eq.~\eqref{EQ:energy_expectation_value} acknowledges 
that the configuration weight may be negative if the sublattice pattern is not fixed.
By way of the identity
\begin{equation}
\frac{\partial w(C)}{\partial h(\tilde{\mathbf{r}})} = \frac{n(\tilde{\mathbf{r}};C)w(C)}{h(\tilde{\mathbf{r}})},
\end{equation}
we find that the downhill direction in the energy landscape {\parameterized} by $\{ h(\tilde{\mathbf{r}}) \}$ is
related to the energy to bond count correlation function
\begin{equation} \label{EQ:energy_bond_corr_func}
G_k(\tilde{\mathbf{r}}) \equiv -\frac{\partial E}{\partial \log h(\tilde{\mathbf{r}})}  = \langle\!\langle H \rangle\!\rangle_k \langle\!\langle n(\tilde{\mathbf{r}}) \rangle\!\rangle_k
- \langle\!\langle H n(\tilde{\mathbf{r}}) \rangle\!\rangle_k.
\end{equation}
In anticipation of Eq.~\eqref{EQ:opt_scheme}, we have used $\langle\!\langle \cdot  \rangle\!\rangle_k$ to denote averaging 
with respect to the $k^{\text{th}}$ Monte Carlo bin.

Our optimization procedure is carried out as follows.
For a given logarithmic amplitude $\lambda^{(1)} = \log h(\tilde{\mathbf{r}})$,
we generate a sequence of (not always energy-reducing) steps
\begin{equation} \label{EQ:opt_scheme}
\lambda^{(k+1)}:= \frac{\lambda^{(k)}R\,\delta \lambda}{k^{1/3}} \sgn G_k.
\end{equation}
$R$ is a random number chosen from the uniform distribution on the interval $[0,1]$,
and $k = 1, 2, \ldots, 1000$ counts the steps taken through the landscape.
The $1/3$ power ensures that the step size envelope decreases by a factor 10 over the course of 1000 steps.
The optimization is run repeatedly with restarts for step sizes beginning at $\delta \lambda = 0.1$ and 
reduced by successive powers of two until convergence is achieved.

The most serious difficulty is that the correlation function estimates $G_k(\tilde{\mathbf{r}})$
become increasingly noisy for large system sizes, to the point
where the determination of $\sgn G_k(\tilde{\mathbf{r}})$ is no longer reliable.
The problem is most acute for the longest bonds, which appear least frequently
and thus have the worst statistics. (The bond amplitudes, which represent the 
probability of a given type of bond appearing during the Monte Carlo sampling, fall
off rapidly as a function of bond length.)

In small amounts, this noise does not interfere with the energy optimization.
It simply overlays a randomizing motion, somewhat akin to the effect of nonzero temperature in simulated annealing.
Nonetheless, good convergence requires that the noise fall below a certain 
threshold (set by the depth and curvature of the well in which the energy minimum sits.)
In practice, mitigating the noise means taking the Monte Carlo bin size large enough so 
that the longest bonds in the system (with length $\lvert \tilde{\mathbf{r}} \rvert \sim L$) appear often enough in the sampling. 
This consideration sets the limit on the systems sizes we can optimize.

\section{\label{SEC:sign_problem}Sign-problematic simulations}

The energy computed by ignoring signs [i.e., by
sampling with respect to the \emph{magnitude} of Eq.~\eqref{EQ:Monte_Carlo_weight}] is
\begin{equation}
E^\star = \frac{\sum_{C} H(C) \lvert w(C)\rvert }{\sum_{C} \lvert w(C)\rvert} \equiv \llbracket H \rrbracket.
\end{equation}
Making the substitution $w = \lvert w\rvert \sgn w$, we can rewrite
Eq.~\eqref{EQ:energy_expectation_value} as the ratio
of averages
\begin{equation}
E = \frac{\sum_{C} H(C) \lvert w(C)\rvert \sgn w(C)}{\sum_{C} \lvert w(C)\rvert \sgn w(C)} \equiv 
\frac{\llbracket H \sgn w \rrbracket}{\llbracket \sgn w \rrbracket};
\end{equation}
hence, the energy discrepancy $\Delta E = E^\star - E$ takes the form of a 
correlation function
\begin{equation}
\Delta E = E^\star - E =
\frac{\llbracket H \rrbracket \cdot \llbracket \sgn w \rrbracket- \llbracket H \sgn w \rrbracket }{ \llbracket \sgn w \rrbracket }.
\end{equation}

If the $\sgn w$ term fluctuates within the simulation so that $\llbracket \sgn w \rrbracket \approx 0$, 
evaluation of $\Delta E$ is impossible due to large statistical uncertainties. Despite this, the actual 
value of $\Delta E$ may itself be small if there is only a weak correlation between the sign and the energy estimator.
Moreover, $\Delta E$ is identically zero if the $h(\mathbf{r})$ values evolve to produce a static
sublattice {\labelling}. So, at the very least, we can view as a rigorous result the fact that no new static pattern emerged
over the course of our simulations.


\begin{thebibliography}{99}

\bibitem{MacDonald88} A.\ H.\ MacDonald, S.\ M.\ Girvin, and D.\ Yoshioka, Phys.\ Rev.\ B {\bf 37}, 9753 (1988); 
Phys.\ Rev.\ B {\bf 41}, 2565 (1990).

\bibitem{Fujimoto05} S.\ Fujimoto, Phys.\ Rev.\ B {\bf 72}, 024429 (2005).
\bibitem{Lauchli05} A.\ L\"{a}uchli, J.\ C.\ Domenge, C.\ Lhuillier, P.\ Sindzingre, and M.\ Troyer,
Phys.\ Rev.\ Lett. {\bf 95}, 137206 (2005).
\bibitem{Sandvik07} A.\ W.\ Sandvik, Phys.\ Rev.\ Lett.\ {\bf 98}, 227202 (2007).
\bibitem{Beach07a} K.\ S.\ D.\ Beach and A.\ W.\ Sandvik, Phys.\ Rev.\ Lett.\ {\bf 99}, 047202 (2007).
\bibitem{Majumdar12} K.\ Majumdar, D.\ Furton, and G.\ S.\ Uhrig, Phys.\ Rev.\ B {\bf 85}, 144420 (2012).

\bibitem{Diep05}``Frustrated spin systems,'' edited by H. T. Diep editor, (World-Scientific, Singapore, 2005). ISBN 978-981-256-091-9
\bibitem{Lacroix11} ``Introduction to Frustrated Magnetism:\ Materials, Experiments, Theory,'' 
edited by C. Lacroix, P. Mendels, and F. Mila (Springer, Berlin, 2011). ISBN 978-3-642-10588-3

\bibitem{Anderson73} P.\ W.\ Anderson, Mater.\ Res.\ Bull.\ {\bf 8}, 153 (1973).
\bibitem{Affleck88} I.\ Affleck and J.\ B.\ Marston, Phys.\ Rev.\ B {\bf 37}, 3774 (1988).
\bibitem{Read89} N.\ Read and S.\ Sachdev, Phys.\ Rev.\ Lett.\ {\bf 62}, 1694 (1989).

\bibitem{Misguich02} G.\ Misguich, C.\ Lhuillier, M.\ Mambrini, and P.\ Sindzingre, Eur.\ Phys.\ J. B {\bf 26}, 167 (2002).
\bibitem{Hastings04} M.\ B.\ Hastings, Phys.\ Rev.\ B {\bf 69}, 104431 (2004).

\bibitem{Marshall55} W.\ Marshall, Proc.\ Roy.\ Soc.\ A {\bf 48}, 232 (1955).

\bibitem{Beard96} B.\ B.\ Beard, U.-J. Wiese, Phys.\ Rev.\ Lett.\ {\bf 77}, 5130 (1996).
\bibitem{Syliuasen02} O.\ F.\ Sylju\aa{}sen and A.\ W.\ Sandvik, Phys.\ Rev.\ E {\bf 66}, 046701 (2002).
\bibitem{Evertz03} H.\ G.\ Evertz, Adv.\ Phys. {\bf 52}, 1 (2003).

\bibitem{Rumer32} G.\ Rumer, Gottingen Nachr.\ Tech.\ {\bf 1932}, 377 (1932).
\bibitem{Pauling33} L.\ Pauling, J.\ Chem.\ Phys.\ {\bf 1}, 280 (1933).
\bibitem{Hulthen38} L.\ Hulth\'{e}n, Ark.\ Mat.\ Atron. Fys.\ {\bf 26a}, 1 (1938).
\bibitem{Fazekas74} P.\ Fazekas and P.\ W.\ Anderson, Philos.\ Mag.\ {\bf 30}, 23 (1974).
\bibitem{Beach06} K.\ S.\ D.\ Beach and A. W.\ Sandvik, Nucl.\ Phys.\ B {\bf 750}, 142 (2006).
\bibitem{Beach08} K.\ S.\ D.\ Beach, M.\ Mambrini, and F.\ Alet, Phys.\ Rev.\ B {\bf 77}, 146401 (2008).

\bibitem{Liang90} S.\ Liang, Phys.\ Rev.\ B {\bf 42}, 6555 (1990); Phys.\ Rev.\ Lett.\ {\bf 64}, 1597 (1990).
\bibitem{Santoro99} G.\ Santoro, S.\ Sorella, L.\ Guidoni, A.\ Parola, and E.\ Tosatti, Phys.\ Rev.\ Lett.\ {\bf 83}, 3065 (1999).
\bibitem{Sandvik05} A.\ W.\ Sandvik, Phys.\ Rev.\ Lett.\ {\bf 95}, 207203 (2005).
\bibitem{Sandvik10} A.\ W.\ Sandvik and H.\ G.\ Evertz, Phys.\ Rev.\ B {\bf 82}, 024407 (2010). 
\bibitem{Banerjee10} A.\ Banerjee and K.\ Damle, J.\ Stat.\ Mech.\ P08017 (2010).

\bibitem{Liang88} S.\ Liang, B.\ Doucot, and P.\ W.\ Anderson, Phys.\ Rev.\ Lett.\ {\bf 61}, 365 (1988).
\bibitem{Lin12} Y.-C.\ Lin, Y.\ Tang, J.\ Lou, A.\ W.\ Sandvik, Phys. Rev. B {\bf 86}, 144405 (2012).

\bibitem{Beach07b} K.\ S.\ D.\ Beach, arxiv:0707.0297.
\bibitem{Hasselmann06} In the more familiar spin-wave language, the justification is the
weak interaction between magnons at long range; see N.\ Hasselmann and P.\ Kopietz, Europhys.\ Lett.\ {\bf 74}, 1067 (2006).

\bibitem{Chandra90} P.\ Chandra, P.\ Coleman, and A.\ I.\ Larkin, Phys.\ Rev.\ Lett.\ {\bf 64}, 88 (1990).

\bibitem{Lou07} J.\ Lou and A.\ W.\ Sandvik, Phys.\ Rev.\ B {\bf 76}, 104432 (2007).
\bibitem{Beach09} K.\ S.\ D.\ Beach, Phys.\ Rev.\ B {\bf 79}, 224431 (2009).

\bibitem{Richter94} J.\ Richter, N.\ B.\ Ivanov, and K.\ Retzlaff, Europhys.\ Lett.\ {\bf 25}, 545 (1994).

\bibitem{Albuquerque10} A.\ F.\ Albuquerque and F.\ Alet, Phys.\ Rev.\ B {\bf 82}, 180408(R) (2010).
\bibitem{Tang11} Y.\ Tang, A.\ W.\ Sandvik, C.\ L. Henley, Phys.\ Rev.\ B {\bf 84}, 174427 (2011).

\bibitem{Moreo90} A.\ Moreo, E.\ Dagotto, Th.\ Jolicoeur, and J.\ Riera, Phys.\ Rev.\ B {\bf 42}, 6283 (1990).
\bibitem{Chubukov91} A.\ Chubukov, Phys.\ Rev.\ B {\bf 44}, 392 (1991).
\bibitem{Ferrer93} J.\ Ferrer, Phys.\ Rev.\ B {\bf 47}, 8769 (1993).
\bibitem{Ceccatto93} H.\ A.\ Ceccatto, C.\ J.\ Gazza, and A.\ E.\ Trumper, Phys.\ Rev.\ B {\bf 47}, 12329 (1993).

\bibitem{Dagotto89} E.\ Dagotto and A.\ Moreo, Phys.\ Rev.\ Lett.\ {\bf 63}, 2148 (1989).
\bibitem{Schulz96} H.\ J.\ Schulz, T.\ A.\ L.\ Ziman, and D.\ Poilblanc, J.\ Phys.\ I France {\bf 6}, 675 (1996).
\bibitem{Oitma96} J.\ Oitmaa and Z.\ Weihong, Phys.\ Rev.\ B {\bf 54}, 3022 (1996).
\bibitem{Bishop98} R.\ F.\ Bishop, D.\ J.\ J.\ Farnell, and J.\ B.\ Parkinson, Phys.\ Rev.\ B {\bf 58}, 6394 (1998).
\bibitem{Singh99} R.\ R.\ P.\ Singh, Z.\ Weihong, C.\ J.\ Hamer, and J.\ Oitmaa, Phys.\ Rev.\ B {\bf 60}, 7278 (1999).
\bibitem{Sushkov01} O.\ P.\ Sushkov, J.\ Oitmaa, and W.\ Zheng, Phys.\ Rev.\ B {\bf 63}, 104420 (2001).

\bibitem{Richter10} J.\ Richter and J.\ Schulenburg, Eur.\ Phys.\ J.\ B {\bf 73}, 117 (2010).
\bibitem{Reuther10} J.\ Reuther and P.\ W\"{o}lfle, Phys.\ Rev.\ B {\bf 81}, 144410 (2010).

\bibitem{Gelfand89} M. P. Gelfand, R. R. P. Singh, and D. A. Huse, Phys.\ Rev.\ B {\bf 40}, 10801 (1989).
\bibitem{Gelfand90} M. P. Gelfand, Phys.\ Rev.\ B {\bf 42}, 8206 (1990).
\bibitem{Singh90} R. R. P. Singh and R. Narayanan, Phys.\ Rev.\ Lett.\ {\bf 65}, 1072 (1990).
\bibitem{Zhitomirsky96} M.\ E.\ Zhitomirsky and K.\ Ueda, Phys.\ Rev.\ B {\bf 54}, 9007 (1996).
\bibitem{Leung96} P.\ W.\ Leung and N.\ W.\ Lam, Phys.\ Rev.\ B {\bf 53}, 2213 (1996).
\bibitem{Kotov99} V. N. Kotov, J. Oitmaa, O. P. Sushkov, and Z. Weihong, Phys.\ Rev.\ B {\bf 60}, 14613 (1999).
\bibitem{Kotov00} V. N. Kotov and O. P. Sushkov, Phys.\ Rev.\ B {\bf 61}, 11820 (2000).
\bibitem{Capriotti00} L. Capriotti and S. Sorella, Phys.\ Rev.\ Lett.\ {\bf 84}, 3173 (2000).
\bibitem{Takao03} K. Takano, Y. Kito, Y. \={O}no, and K. Sano, Phys.\ Rev.\ Lett.\ {\bf 91}, 197202 (2003).
\bibitem{Mambrini06} M. Mambrini, A. L\"{a}uchli, D. Poilblanc, and F. Mila, Phys.\ Rev.\ B {\bf 74}, 144422 (2006).
\bibitem{Murg09} V.\ Murg, F.\ Verstraete, and J.\ I.\ Cirac, Phys.\ Rev.\ B {\bf 79}, 195119 (2009).
\bibitem{Reuther11} J.\ Reuther, P.\ W\"{o}lfle, R.\ Darradi, W.\ Brenig, M.\ Arlego, and J.\ Richter,
Phys.\ Rev.\ B {\bf 83}, 064416 (2011).
\bibitem{Yu12} J.-F.\ Yu and Y.-J.\ Kao, Phys.\ Rev.\ B {\bf 85}, 094407 (2012).

\bibitem{Capriotti01} L.\ Capriotti, F.\ Becca, A.\ Parola, and S.\ Sorella, Phys.\ Rev.\ Lett.\ {\bf 87}, 097201 (2001).
\bibitem{Chandra88} P.\ Chandra and B.\ Doucot, Phys.\ Rev.\ B {\bf 38}, 9335 (1988).
\bibitem{Figueirido90} F.\ Figueirido, A.\ Karlhede, S.\ Kivelson, S.\ Sondhi, M.\ Rocek, and D.\ S.\ Rokhsar, Phys.\ Rev.\ B {\bf 41}, 4619 (1990).
\bibitem{Oguchi90} T.\ Oguchi and H.\ Kitatani, J.\ Phys.\ Soc.\ Jpn.\ {\bf 59}, 3322 (1990).
\bibitem{Locher90} P.\ Locher, Phys.\ Rev.\ B {\bf 41}, 2537 (1990).
\bibitem{Schulz92} H.\ J.\ Schulz and T.\ A.\ L.\ Ziman, Europhys.\ Lett.\ {\bf 18}, 355 (1992).
\bibitem{Zhong93} Q.\ F.\ Zhong and S.\ Sorella, Europhys.\ Lett.\ {\bf 21}, 629 (1993).
\bibitem{Zhang03} G.-M.\ Zhang, H.\ Hu, and L.\ Yu, Phys.\ Rev.\ Lett.\ {\bf 91}, 067201 (2003).
\bibitem{Capriotti03} L.\ Capriotti, F.\ Becca, A.\ Parola, and S.\ Sorella, Phys.\ Rev.\ B {\bf 67}, 212402 (2003).
\bibitem{Captiotti04a} L.\ Capriotti, D.\ J.\ Scalapino, and S.\ R.\ White, Phys.\ Rev.\ Lett.\ {\bf 93}, 177004 (2004).
\bibitem{Capriotti04b} L.\ Capriotti and S.\ Sachdev, Phys.\ Rev.\ Lett.\ {\bf 93}, 257206 (2004).

\bibitem{Wang11} L.\ Wang, Z.-C.\ Gu, X.-G.\ Wen, and F.\ Verstraete, arXiv:1112.3331v2.
\bibitem{Jiang11} H.-C.\ Jiang, H.\ Yao, and L.\ Balents, Phys.\ Rev.\ B {\bf 86}, 024424 (2012).
\bibitem{Mezzacapo12} F.\ Mezzacapo, Phys.\ Rev.\ B {\bf 86}, 045115 (2012).
\bibitem{Sandvik12} A.\ W.\ Sandvik, Phys.\ Rev.\ B {\bf 85} 134407 (2012).

\bibitem{Block12} M.\ S.\ Block and R.\ K.\ Kaul, Phys.\ Rev.\ B {\bf 86}, 134408 (2012).
\bibitem{Kaul12} R.\ K.\ Kaul, R.\ G.\ Melko, and A.\ W.\ Sandvik, Annu.\ Rev.\ Condens.\ Matter
Phys.\ {\bf 4}, 8.1--8.37 (2013).

\bibitem{Loh90} E.\ Y.\ Loh Jr., J.\ E.\ Gubernatis, R.\ T.\ Scalettar, S.\ R.\ White, D.\ J.\ Scalapino,
and R.\ L.\ Sugar, Phys.\ Rev.\ B {\bf 41}, 9301 (1990).

\bibitem{Richter04} J.\ Richter, J.\ Schulenburg, A.\ Honecker, and D.\ Schmalfu\ss{}, Phys.\ Rev.\ B {\bf 70}, 174454 (2004).
\bibitem{Nakano11} H.\ Nakano and T.\ Sakai, J.\ Phys.\ Soc.\ Jpn.\ {\bf 80}, 053704 (2011).
\bibitem{Lauchli11} A.\ M.\ L\"{a}uchli, J.\ Sudan, and E.\ S.\ S{\o}rensen, Phys.\ Rev.\ B {\bf 83}, 212401 (2011).
\bibitem{Lauchli12} A.\ M.\ L\"{a}uchli and R.\ Johanni, Bull.\ Am.\ Phys.\ Soc.\ {\bf 57}, 1 (2012).
[\url{http://meetings.aps.org/link/BAPS.2012.MAR.H8.7}]

\bibitem{Anderson87} P.\ W.\ Anderson, Science {\bf 235} 1196 (1987).
\bibitem{Yunoki04} S.\ Yunoki and S.\ Sorella, Phys.\ Rev.\ Lett.\ {\bf 92}, 157003 (2004).
\bibitem{Cano10} J.\ Cano and P.\ Fendley, Phys.\ Rev.\ Lett.\ {\bf 105}, 067205 (2010).
\bibitem{Li12} T.\ Li, F.\ Becca, W.\ Hu, and S.\ Sorella, Phys.\ Rev.\ B {\bf 86}, 075111 (2012).

\bibitem{Prokofev01} N.\ Prokof'ev and B.\ Svistunov, Phys.\ Rev.\ Lett.\ {\bf 87}, 160601 (2001).
\bibitem{Sandvik06} A.\ W.\ Sandvik and R.\ Moessner, Phys.\ Rev.\ B {\bf 73}, 144504 (2006).

\bibitem{Lehmer60} D.\ H.\ Lehmer, Proc.\ Sympos.\ Appl.\ Math.\ Combinatorial Analysis, Amer.\ Math.\ Soc.\ {\bf 10}, 179 (1960).
\bibitem{Knuth73} D.\ E.\ Knuth, ``Volume 3: Sorting and Searching,'' The Art of Computer Programming, Addison-Wesley, p.\ 12, (1973). ISBN 0-201-89685-0

\bibitem{Wang13} L.\ Wang, D.\ Poilblac, Z.-C.\ Gu, X.-G.\ Wen, and F.\ Verstraete, arXiv:1112.3331v2.

\bibitem{Havilio99} M.\ Havilio and A.\ Auerbach, Phys.\ Rev.\ Lett.\ {\bf 83}, 4848 (1999).
\bibitem{Havilio00} M.\ Havilio and A.\ Auerbach, Phys.\ Rev.\ B {\bf 62}, 324 (2000).

\end{thebibliography}
\end{document}